\documentclass[twocolumn]{autart}

\usepackage{graphicx}          

\usepackage{amsmath,amssymb,amsfonts}
\usepackage{mathtools}

\usepackage{algorithm, algpseudocode}
\usepackage{eqparbox}
\usepackage{wrapfig}
\usepackage{needspace}

\usepackage[authoryear,round]{natbib}

\usepackage[pageanchor=false,pdfencoding=pdfdoc]{hyperref}
\usepackage{subcaption}

\begin{document}

\begin{frontmatter}

	\title{Synthesizing Neural Network Controllers with Closed-Loop Dissipativity Guarantees}

	\thanks[footnoteinfo]{
		This work was supported in part by the National Science Foundation award CNS-2111688 and by the Air Force Office of Scientific Research grants FA9550-21-1-0288 and FA9550-23-1-0732.
	}

	\author[Berkeley]{Neelay Junnarkar}\ead{neelay.junnarkar@berkeley.edu},    
	\author[Berkeley]{Murat Arcak}\ead{arcak@berkeley.edu},  
	\author[Michigan]{Peter Seiler}\ead{pseiler@umich.edu}               

	\address[Berkeley]{Department of Electrical Engineering and Computer Sciences, University of California, Berkeley}  
	\address[Michigan]{Department of Electrical Engineering and Computer Science, University of Michigan, Ann Arbor}             


	\begin{abstract}                          
		This paper presents a method to synthesize neural network controllers to maximize reward subject to the hard constraint that the feedback system of plant and controller be dissipative, certifying requirements such as stability and $L_2$ gain bounds.
		It considers nonlinear and uncertain plants, modeled as the interconnection of a
		linear time-invariant (LTI) system and an uncertainty block, which incorporates nonlinearities.
		The uncertainty of the plant and the activation functions of the neural network are both described using integral quadratic constraints (IQCs).
		First, a dissipativity condition is derived for uncertain LTI systems.
		Second, this condition is used to construct a linear matrix inequality (LMI) which can be used to synthesize neural network controllers.
		Finally, this convex condition is used in a projection-based training method to synthesize neural network controllers with dissipativity guarantees.
		Numerical examples on an inverted pendulum and a flexible rod on a cart are provided to demonstrate the effectiveness of this approach.
	\end{abstract}

\end{frontmatter}

\section{Introduction}
\label{sec:introduction}

Neural networks have seen recent success in control tasks, particularly through reinforcement learning, due to their ability to express complex nonlinear behavior.
Neural network controllers have been applied to control problems such as cart pole swing-up, walking, and moving multiple degree-of-freedom arms \citep{LillicrapHPHETS15}.
However, neural network controllers trained through standard reinforcement learning techniques suffer from sensitivity to inputs and may display unexpected behavior.
Concerns about neural network controllers are accentuated in safety-critical applications.
This has motivated research into safe neural network control, which focuses on acquiring formal guarantees for neural networks used in control applications, while still leveraging the expressiveness of neural networks.

One class of methods focuses on verifying properties of closed-loop systems involving neural networks.
These methods make no assumptions on the methods used to synthesize the neural network.
\citet{yinStabilityAnalysisUsing2022}, \citet{hashemiCertifyingIncrementalQuadratic2021}, and \citet{pauliLinearSystemsNeural2021} present methods to verify asymptotic stability or find an inner approximation to the region of attraction of a feedback system involving neural networks.
Properties of dynamical systems themselves modeled by neural networks are also analyzed: \citet{barabanovStabilityAnalysisDiscretetime2002} use linear matrix inequality techniques to analyze stability of an equilibrium of a recurrent neural network, and \citet{karnyRecurrentNeuralNetworks1998} analyzes approximation capabilities, controllability, and observability of recurrent neural networks.
These results can be used for verifying closed-loop properties of a neural network controller by modeling the interconnection of the controller and plant as a neural network.
Methods for verifying properties of closed-loops involving neural network controllers are not necessarily amenable to neural network controller synthesis due to the structure imposed by the fixed plant parameters.

A second class of methods focuses on jointly synthesizing neural network controllers and certifying closed-loop properties.
Common techniques include enforcing linear matrix inequality (LMI) constraints through projection \citep{gu_recurrent_2021}, \citep{junnarkar_synthesis_2022} and using the Youla-Kucera parameterization to augment a base controller with a neural network controller \citep{wang_learning_2023}, \citep{barbaraLearningContractingLipschitz2023}, \citep{lawrenceStabilizingReinforcementLearning2024}.
Additional methods include \citet{jinStabilityCertifiedReinforcementLearning2020}, who restrict the partial derivatives of the output of the neural network with respect to its input; and \citet{zhouNeuralLyapunovControl2024}, who use satisfiability module theories to guarantee closed-loop stability.

The synthesis methods above differ in controller model and plant class.
Controller models range from recurrent neural networks \citep{gu_recurrent_2021}, \citep{damicoDatabasedControlDesign2023} to memoryless feedforward neural networks \citep{jinStabilityCertifiedReinforcementLearning2020}, \citep{zhouNeuralLyapunovControl2024}, \citep{lawrenceStabilizingReinforcementLearning2024} to models leveraging implicit neural networks \citep{junnarkar_synthesis_2022}, \citep{wang_learning_2023}, \citep{barbaraLearningContractingLipschitz2023}, which encompass both of the preceding.
Some methods are restricted to linear time-invariant (LTI) plants \citep{wang_learning_2023}, \citep{lawrenceStabilizingReinforcementLearning2024}, while others handle nonlinear or uncertain plants \citep{barbaraLearningContractingLipschitz2023}, \citep{zhouNeuralLyapunovControl2024}, with a standard technique being to decompose the plant into an LTI part and a nonlinear part abstracted with integral quadratic constraints (IQCs) or Lipschitz bounds \citep{gu_recurrent_2021}, \citep{junnarkar_synthesis_2022}, \citep{jinStabilityCertifiedReinforcementLearning2020}.

An important consideration in safe neural network control is computational tractability of controller synthesis methods.
A method used in \citep{gu_recurrent_2021} and \citep{junnarkar_synthesis_2022} is to project the controller parameters to a safe set by solving a semidefinite program (SDP) after each reinforcement learning parameter update step.
\citet{pauli_barrier_2022} show that barrier function methods can be used to convert training under SDP constraints into an unconstrained training problem.
\citet{wang_learning_2023} and \citet{barbaraLearningContractingLipschitz2023} use unconstrained gradient-descent methods.

In this paper, we present a method to synthesize neural network controllers for a class of nonlinear and uncertain plants to guarantee closed-loop dissipativity and maximize reward.
The closed-loop dissipativity specification enables, for example, certifying stability, $L_2$ gain bounds for robustness to disturbances, and passivity for use in interconnected systems.
The reward function is used to express auxiliary performance metrics or otherwise shape controller behavior.
Closed-loop dissipativity is a hard requirement while maximizing reward is soft.
We model the plant as the interconnection of an LTI system and an uncertainty and nonlinearity.
The uncertainty can be used to describe unmodeled dynamics that result from the use of reduced-order models. 
Since we model both nonlinearities and uncertainties with integral quadratic constraints (IQCs) \citep{megretski_system_1997}, we use the word `uncertainty' to encompass `nonlinearity'.

Control of plants described by the interconnection of an LTI system and an uncertainty has been widely explored using IQCs to describe the uncertainty and linear matrix inequality (LMI) optimization for computation \citep{boydLinearMatrixInequalities1994}.
For example, \citet{veenmanIQCsynthesisGeneralDynamic2014} address $\mathcal{H}_\infty$ control of uncertain LTI systems and \citet{wang_seiler_lpv_2016} further consider $\mathcal{H}_\infty$ control of uncertain linear parameter varying systems, both using LTI controllers.
\citet{veenmanRobustStabilityPerformance2016} give a comprehensive overview of analysis of uncertain LTI systems using IQCs and LMIs, with examples focused on $\mathcal{H}_\infty$ control.

To model the neural network controller, we leverage the implicit neural network (also known as equilibrium network), a neural network model which encompasses common neural network architectures, including the fully connected feedforward network \citep{el_ghaoui_implicit_2021}.
Conveniently for analysis, a controller with memory based on the implicit neural network can be modeled as an uncertain LTI system where the ``uncertainty'' consists of the activation functions.

In this paper, we provide sufficient conditions for an uncertain LTI system to be dissipative when the uncertainty is described by integral quadratic constraints.
We leverage these conditions in deriving sufficient convex conditions for the feedback system of a neural network controller and an uncertain LTI system to be dissipative.
We present a procedure to train a neural network controller to maximize a reward function while guaranteeing a dissipativity condition.
We demonstrate the benefits of our proposed controller in simulated examples on an inverted pendulum and on a flexible rod on a cart.
Our primary contribution is that we simultaneously guarantee robustness to exogenous disturbances, which we achieve through the unifying framework of dissipativity, and robustness to model uncertainty.

Compared to \citet{gu_recurrent_2021}, who consider plant uncertainty and stability, we use a significantly larger class of dynamical systems to model the controller, and also consider robustness to disturbances through the framework of dissipativity.
Compared to \citet{jinStabilityCertifiedReinforcementLearning2020}, who consider plant uncertainty and stability through state-feedback with a memoryless neural network controller, we use a larger class of neural network controllers with dynamics, output-feedback, robustness to disturbances, and a less conservative characterization of the neural network.
Compared to \citet{junnarkar_synthesis_2022}, who consider stability of plants with sector-bounded nonlinearities, we use a more flexible characterization of nonlinearities and uncertainties, and robustness to disturbances.
Compared to \citet{wang_learning_2023} and \citet{barbaraLearningContractingLipschitz2023}, who consider robustness to disturbances with LTI or nonlinear plants, we consider a broader class of plants and uncertainties, and unlike \citet{barbaraLearningContractingLipschitz2023}, we do not assume the existence of a stabilizing base controller.

The rest of this paper is structured as follows.
In Section~\ref{sec:problem-setup}, the plant and controller models are presented, and characterizations of uncertainty and performance requirements are given.
In Section~\ref{sec:uncertain-lti-system}, a condition for verifying dissipativity is presented.
In Section~\ref{sec:controller-synthesis}, a convex condition equivalent to the one in Section~\ref{sec:uncertain-lti-system} is derived, and a procedure for synthesizing a controller using reinforcement learning and linear matrix inequalities is presented.
In Section~\ref{sec:experiments}, the controller synthesis procedure is demonstrated in simulated examples on an inverted pendulum and a flexible rod on a cart.

\subsection{Notation}
We let the symbol $\mathbb{R}_{\geq 0}$ denote the set of nonnegative real numbers, $\mathcal{L}_2^n$ the space of square-integrable functions from $\mathbb{R}_{\geq 0}$ to $\mathbb{R}^n$, and  $\mathcal{L}_{2e}^n$ the space of functions from  $\mathbb{R}_{\geq 0}$ to $\mathbb{R}^n$ which are square-integrable on the interval $[0, T]$ for all $T > 0$.
We drop the superscript denoting the dimension of the codomain if it is clear from context.
We use bold symbols to denote signals and their non-bold counterparts to denote points.
For example, $\boldsymbol{x_p}$ denotes the element of $\mathcal{L}_{2e}^{n_p}$ and $x_p$ denotes an element of $\mathbb{R}^{n_p}$.
For brevity, we use $(\star)^\top X y$ to denote $y^\top X y$ when the expression that appears in place of $y$ is lengthy.

\section{Problem Setup}\label{sec:problem-setup}
We consider the problem of controlling a plant with a neural network controller to both satisfy a performance requirement and maximize a reward.
We consider a plant \textit{design model} of the form of an uncertain linear time-invariant system.
The model/environment with respect to which we maximize reward need not be the same as the design model (it can, for example, be a higher-fidelity simulation model).
This is a constrained optimization problem where the performance requirement is the constraint and the reward is the optimization objective.

\begin{figure}[tb]
	\centering
	\includegraphics[scale=0.8]{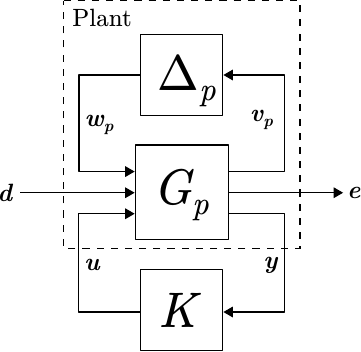}
	\caption{Interconnection of plant, formed of LTI system $G_p$ and uncertainty $\Delta_p$, and controller $K$.}
	\label{fig:Gp-Deltap-K}
\end{figure}

The uncertain LTI plant design model is modeled by the interconnection of an LTI system and an uncertainty $\Delta_p$. This is written as
\begin{equation} \label{eq:plant} 
	\begin{aligned}
		\begin{bmatrix} \dot{\boldsymbol{x}}_{\boldsymbol{p}}(t) \\ \boldsymbol{v_p}(t) \\ \boldsymbol{e}(t) \\ \boldsymbol{y}(t) \end{bmatrix}
		                    & =
		\begin{bmatrix}
			A_p     & B_{p w}   & B_{p d}   & B_{p u}   \\
			C_{p v} & D_{p v w} & D_{p v d} & D_{p v u} \\
			C_{p e} & D_{p e w} & D_{p e d} & D_{p e u} \\
			C_{p y} & D_{p y w} & D_{p y d} & 0
		\end{bmatrix}
		\begin{bmatrix} \boldsymbol{x_p}(t) \\ \boldsymbol{w_p}(t) \\ \boldsymbol{d}(t) \\ \boldsymbol{u}(t) \end{bmatrix}, \\
		\boldsymbol{w_p}(t) & = \Delta_p(\boldsymbol{v_p})(t),
	\end{aligned}
\end{equation}
where $x_p \in \mathbb{R}^{n_p}$ is the plant state, $d \in \mathbb{R}^{n_d}$ is an exogenous disturbance, $e \in \mathbb{R}^{n_e}$ is the performance output, $u \in \mathbb{R}^{n_u}$ is the control input, $y \in \mathbb{R}^{n_y}$ is the measurement output, $\boldsymbol{v_p} \in \mathcal{L}_{2e}^{n_v}$ is the input to the uncertainty $\Delta_p$, and $\boldsymbol{w_p} \in \mathcal{L}_{2e}^{n_w}$ is the output of the uncertainty.
We assume this system is \textit{well-posed}, meaning that there exists a unique trajectory $(\boldsymbol{x_p}, \boldsymbol{v_p}, \boldsymbol{w_p}, \boldsymbol{e}, \boldsymbol{y})$, with each signal in $\mathcal{L}_{2e}$, satisfying \eqref{eq:plant} for each combination of plant initial condition $\boldsymbol{x_p}(0) = x_{p 0}$, input $\boldsymbol{u} \in \mathcal{L}_{2e}^{n_u}$, and disturbance $\boldsymbol{d}\in \mathcal{L}_{2e}^{n_d}$.
The interconnection of the plant and controller is depicted in Figure~\ref{fig:Gp-Deltap-K}.

We use the typical form of the reward function used in reinforcement learning \citep{suttonReinforcementLearningIntroduction2020}.
The reward over a time horizon $T$, under a particular control law, is given by the integral of a reward function $r: \mathbb{R}^{n_p} \times \mathbb{R}^{n_u} \to \mathbb{R}$:
\begin{equation}
	\mathbb{E}\left[\int_{0}^T r(\boldsymbol{x_p}(t), \boldsymbol{u}(t)) dt\right],
\end{equation}
where the expectation is taken over a distribution of disturbance signals and initial plant states.
Note that, even when the plant is LTI, a non-quadratic reward function may lead to a nonlinear optimal controller.

\subsection{Neural Network Controller Model}

\begin{figure}[tb]
	\centering
	\includegraphics[scale=0.8]{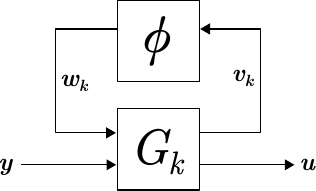}
	\caption{The neural network controller is modeled as the interconnection of an LTI system $G_k$ and an activation function $\phi$. The function $\phi$ represents the scalar activation functions of the neural network stacked together }
	\label{fig:K}
\end{figure}

The neural network controller is modeled as the interconnection of an LTI system and activation functions $\phi$ (see Figure~\ref{fig:K}):
\begin{equation} \label{eq:controller} 
	\begin{aligned}
		\begin{bmatrix} \dot{\boldsymbol{x}}_{\boldsymbol{k}}(t) \\ \boldsymbol{v_k}(t) \\ \boldsymbol{u}(t) \end{bmatrix}
		                    & =
		\begin{bmatrix}
			A_k     & B_{k w}   & B_{k y}   \\
			C_{k v} & D_{k v w} & D_{k v y} \\
			C_{k u} & D_{k u w} & D_{k u y}
		\end{bmatrix}
		\begin{bmatrix}
			\boldsymbol{x_k}(t) \\ \boldsymbol{w_k}(t) \\ \boldsymbol{y}(t)
		\end{bmatrix}, \\
		\boldsymbol{w_k}(t) & = \phi(\boldsymbol{v_k}(t)),
	\end{aligned}
\end{equation}
where $x_k \in \mathbb{R}^{n_k}$ is the controller state, $v_k \in \mathbb{R}^{n_\phi}$ is the input to the activation function $\phi$, and $w_k \in \mathbb{R}^{n_\phi}$ is the output of the activation function $\phi$.
Many neural networks can be modeled in this form, which mimics the form of the plant model.

The activation function $\phi$ is memoryless and applied elementwise: $\phi$ is constructed from scalar activation functions $\phi_1, \dots, \phi_{n_\phi}$ and $w_{k, i} = \phi_i(v_{k, i})$.
Each scalar activation function is sector-bounded in $[0, 1]$, and slope-restricted in $[0, 1]$.
Common activation functions that satisfy the sector bounds and slope restrictions are tanh and ReLU.

The controller \eqref{eq:controller} is well-posed if there exists a unique solution for $w_k$ in the implicit equation $w_k = \phi(C_{k v} x_k + D_{k v w} w_k + D_{k v y} y_k)$ for all $x_k$ and $y_k$.
Ensuring well-posedness requires constraining $D_{k v w}$.
Many sufficient conditions are available for well-posedness of this equation (see, for example, \citep{el_ghaoui_implicit_2021}).
We discuss a condition particularly relevant to this work in Section~\ref{subsec:training}, which strengthens a positive semidefinite condition in Sections~\ref{sec:uncertain-lti-system}~and~\ref{sec:controller-synthesis} to a positive definite condition.
Note that neural network architectures where the output is explicitly computed make these equations well-posed, even when modeled in the implicit form.

With $A_k$, $B_{k w}$, $B_{k y}$, $C_{k v}$, and $C_{k u}$ equal to zero, this controller model simplifies to $\boldsymbol{u}(t) = D_{k u w} \boldsymbol{w_k}(t) + D_{k u y} \boldsymbol{y}(t)$ where $\boldsymbol{w_k}(t) = \phi(D_{k v w}\boldsymbol{w_k}(t) + D_{k v y} \boldsymbol{y}(t))$.
This is a static implicit neural network (INN).
An analysis of the INN can be found in \citep{el_ghaoui_implicit_2021} and \citep{bai_deep_2019}.
Common classes of neural networks, including fully connected feedforward neural networks, convolutional layers, max-pooling layers, and residual networks, can be put into the form of an INN \citep{el_ghaoui_implicit_2021}.
The following example models a feedforward neural network in the form of an implicit neural network.

\begin{exmp}
	Consider the $L$-layer feedforward neural network with input $y$ and output $u$:
	\begin{align*}
		u = W_L w_L + b_L, w_{l+1} = \phi_l(W_l w_l + b_l), w_0 = y.
	\end{align*}
	The vectors $w_1, \dots, w_L$ are the values of the hidden layers, $W_0, \dots, W_L$ are the weight matrices of the neural network, $b_0, \dots, b_L$ are the biases, and $\phi_0, \dots, \phi_{L-1}$ are the scalar activation functions of the layer stacked together.
	With $w_k = \begin{bmatrix} w_1^\top & \cdots & w_L^\top \end{bmatrix}^\top$ and $\phi = \begin{bmatrix} \phi_0^\top & \cdots & \phi_{L-1}^\top \end{bmatrix}^\top$, this can be represented by the following INN:
	\begin{align*}
		u & = \underbrace{\begin{bmatrix} 0 & \cdots & 0 & W_l \end{bmatrix}}_{D_{kuw}} w_k + \underbrace{\begin{bmatrix} 0 & b_L \end{bmatrix}}_{D_{kuy}} \begin{bmatrix} y \\ 1 \end{bmatrix} \\
		w & = \phi\left(
		\underbrace{\begin{bsmallmatrix}
				            0 & \cdots & & & 0 \\
				            W_1 & 0 & \cdots & &  \\
				            0 & W_2 & 0 & \cdots \\
				            \vdots & \ddots & \ddots & \ddots & \vdots \\
				            0 & \cdots & 0 & W_{L-1} & 0
			            \end{bsmallmatrix}}_{D_{kvw}} w_k
		+ \underbrace{\begin{bsmallmatrix}
				              W_0 & b_0 \\
				              0 & b_1 \\
				              \vdots & \vdots \\
				              0 & b_{L-1}
			              \end{bsmallmatrix}}_{D_{kvy}} \begin{bmatrix} y \\ 1 \end{bmatrix}
		\right).
	\end{align*}
\end{exmp}

The augmentation of the INN with dynamics and controller state $x_k$ makes this a recurrent implicit neural network (RINN).
Controllers with memory are useful in applications with partially-observed plants \citep{callier_linear_1991}.
Modeling the neural network controller similarly to the plant simplifies analysis.
This model, or similar, has been used for control in \citep{junnarkar_synthesis_2022}, and \citep{wang_learning_2023}.
With $D_{k v w} = 0$, this controller simplifies to a recurrent neural network similar to the one used in \citep{gu_recurrent_2021}.

\subsection{Dissipativity}
We model the control performance requirement with the notion of dissipativity.
Under assumptions for existence and uniqueness of solutions, the system
\begin{equation} \label{eq:general-dynamics}
	\begin{alignedat}{2}
		\dot{\boldsymbol{x}}(t) = f(\boldsymbol{x}(t), \boldsymbol{d}(t)) & \quad & \boldsymbol{e}(t) = g(\boldsymbol{x}(t), \boldsymbol{d}(t)),
	\end{alignedat}
\end{equation}
where $x \in \mathbb{R}^n$, is dissipative with respect to a supply rate $s: \mathbb{R}^{n_d} \times \mathbb{R}^{n_e} \to \mathbb{R}$ if there exists a nonnegative storage function $S: \mathbb{R}^{n} \to \mathbb{R}_{\geq 0}$ such that
\begin{equation*}
	S(\boldsymbol{x}(T)) - S(\boldsymbol{x}(0)) \leq \int_{0}^{T} s(\boldsymbol{d}(t), \boldsymbol{e}(t)) dt
\end{equation*}
for any $T \geq 0$ and trajectory $(\boldsymbol{x}, \boldsymbol{d}, \boldsymbol{e})$ that satisfies the dynamics in \eqref{eq:general-dynamics}.
A thorough presentation of dissipativity can be found in \citep{willems_dissipative1_1972} and \citep{willems_dissipative2_1972}.

In particular, we consider a quadratic supply rate
\begin{align} \label{eq:supply-rate}
	s(d, e)  = \begin{bmatrix}d \\ e\end{bmatrix}^\top X \begin{bmatrix} d \\ e \end{bmatrix}
\end{align}
where
\begin{equation*}
	X =  \begin{bmatrix} X_{d d} & X_{d e} \\ X_{d e}^\top & X_{e e} \end{bmatrix}
\end{equation*}
and a quadratic storage function
\begin{equation}
	S(x) = x^\top P x
\end{equation}
where $P \succeq 0$.
Important classes of quadratic supply rates include $s(d, e) = 0$, which corresponds to stability (when we also impose $P \succ 0$), $s(d, e) = \gamma^2 \|d\|^2 - \|e\|^2$, which corresponds to an $L_2$ gain bound of $\gamma$, and $s(d, e) = d^\top e$, which corresponds to passivity.

\subsection{Integral Quadratic Constraints}

The block $\Delta_p$ may represent a variety of uncertainties, including unmodeled dynamics and uncertain parameters, and can also be used to capture parts of dynamics that are known but difficult to analyze, such as nonlinearities.
To characterize $\Delta_p$ in a manner convenient for dissipativity analysis, we describe the relationship between its inputs and outputs using a form of time-domain integral quadratic constraints (IQCs) \citep{megretski_system_1997} \citep{schererDissipativityIntegralQuadratic2022} \citep{seiler_stability_2015} \citep{veenmanIQCsynthesisGeneralDynamic2014}.
A library of uncertainty classes and IQCs to characterize them is presented in \citep{megretski_system_1997}.

Define a stable filter $\Psi_p$ by
\begin{equation} \label{eq:plant-filter}
	\begin{aligned}
		\begin{bmatrix}
			\dot{\boldsymbol{\psi}}_{\boldsymbol{p}}(t) \\ \boldsymbol{z_p}(t)
		\end{bmatrix}
		=
		\begin{bmatrix}
			A_\psi & B_{\psi v} & B_{\psi w} \\
			C_\psi & D_{\psi v} & D_{\psi w}
		\end{bmatrix}
		\begin{bmatrix}
			\boldsymbol{\psi_p}(t) \\ \boldsymbol{v_p}(t) \\ \boldsymbol{w_p}(t)
		\end{bmatrix}
	\end{aligned}
\end{equation}
where $\boldsymbol{\psi_p}(0) = 0$. Then, with $M_{\Delta_p} = M_{\Delta_p}^\top$, we say $\Delta_p$ satisfies the IQC defined by $(\Psi_p, M_{\Delta_p})$ if
\begin{equation}
	\int_0^T \boldsymbol{z_p}(t)^\top M_{\Delta_p} \boldsymbol{z_p}(t) dt \geq 0 \quad \forall T \geq 0
\end{equation}
for all $\boldsymbol{v_p}$ and  $\boldsymbol{w_p} = \Delta_p(\boldsymbol{v_p})$.
This is often referred to as a ``hard'' IQC \citep{megretski_system_1997}.
In this paper, we only use hard IQCs, so we henceforth refer to them just as IQCs.

As a special case, we say $\Delta_p$ satisfies a static IQC defined by $M_{\Delta_p}$ if $\Delta_p$ satisfies the IQC defined by $(I, M_{\Delta_p})$.
This is the case where $\boldsymbol{z_p}(t) = \begin{bmatrix} \boldsymbol{v_p}(t)^\top & \boldsymbol{w_p}(t)^\top \end{bmatrix}^\top$. As a further special case, we say $\Delta_p$ satisfies the quadratic constraint (QC) defined by $M_{\Delta_p}$ if
\begin{equation*}
	\begin{bmatrix} \boldsymbol{v_p}(t) \\ \boldsymbol{w_p}(t) \end{bmatrix}^\top
	M_{\Delta_p}
	\begin{bmatrix} \boldsymbol{v_p}(t) \\ \boldsymbol{w_p}(t) \end{bmatrix}
	\geq 0 \quad \forall t \geq 0.
\end{equation*}
We say dynamic IQC to refer to IQCs with filters when it is not clear from context if a special case of IQC is being referred to.

The controller activation function $\phi$, due to each scalar activation function being sector-bounded in $[0, 1]$, satisfies the quadratic constraint defined by
\begin{equation}
	M_\phi = \begin{bmatrix}
		0 & \Lambda \\ \Lambda & -2 \Lambda
	\end{bmatrix}
\end{equation}
where $\Lambda \succeq 0$ and diagonal \citep{fazlyab_safety_2022}.

\section{Dissipativity Certification}
\label{sec:uncertain-lti-system}

In this section, we derive conditions for an uncertain LTI system to be dissipative with respect to the specified supply rate.

Let the uncertain LTI system be:
\begin{equation}\label{eq:uncertain-lti-sys} 
	\begin{aligned}
		\begin{bmatrix}
			\dot{\boldsymbol{x}}(t) \\ \boldsymbol{v}(t) \\ \boldsymbol{e}(t)
		\end{bmatrix}
		                  & =
		\begin{bmatrix}
			A     & B_{w}   & B_{d}   \\
			C_{v} & D_{v w} & D_{v d} \\
			C_{e} & D_{e w} & D_{e d}
		\end{bmatrix}
		\begin{bmatrix}
			\boldsymbol{x}(t) \\ \boldsymbol{w}(t) \\ \boldsymbol{d}(t)
		\end{bmatrix} \\
		\boldsymbol{w}(t) & = \Delta(\boldsymbol{v})(t),
	\end{aligned}
\end{equation}
where $x \in \mathbb{R}^n$, $v \in \mathbb{R}^{n_v}$, $w \in \mathbb{R}^{n_w}$, $d \in \mathbb{R}^{n_d}$, and $e \in \mathbb{R}^{n_e}$.
Assume this system is well-posed.

\subsection{Static IQCs}\label{subsec:static-iqc}

We begin with the simplified case where the uncertainty $\Delta$ satisfies a static IQC defined by $M$, and generalize this in the next subsection.

\begin{lem} \label{lemma:static-iqc-dissipativity}
	The system defined in \eqref{eq:uncertain-lti-sys}, with the uncertainty satisfying a static IQC defined by $M$, is dissipative with respect to the quadratic supply rate in \eqref{eq:supply-rate} if there exist $P \succeq 0$ and $\lambda \geq 0$ such that
	\begin{equation}\label{eq:open-loop-dissipativity}
		\begin{gathered}
			\begin{bmatrix}
				A^\top P + P A & P B_w & P B_d \\
				B_w^\top P     & 0     & 0     \\
				B_d^\top P     & 0     & 0
			\end{bmatrix}
			+ \lambda (\star)^\top M \begin{bmatrix}
				C_v & D_{v w} & D_{v d} \\
				0   & I       & 0
			\end{bmatrix} \\
			- (\star)^\top X \begin{bmatrix}
				0   & 0       & I       \\
				C_e & D_{e w} & D_{e d}
			\end{bmatrix}
			\preceq 0
		\end{gathered}
	\end{equation}
\end{lem}
\begin{pf}
	Consider a trajectory of the system. Multiply \eqref{eq:open-loop-dissipativity} on the left and right by $\begin{bmatrix} \boldsymbol{x}(t)^\top & \boldsymbol{w}(t)^\top & \boldsymbol{d}(t)^\top \end{bmatrix}^\top$ and its transpose.
	Note that the first term is then equal to $\boldsymbol{x}(t)^\top P \dot{\boldsymbol{x}}(t) + \dot{\boldsymbol{x}}(t)^\top P \boldsymbol{x}(t)$, the second term is equal to $\lambda (\star)^\top M \begin{bmatrix} \boldsymbol{v}(t)^\top & \boldsymbol{w}(t)^\top \end{bmatrix}^\top$, and the third term is equal to $-(\star)^\top X \begin{bmatrix} \boldsymbol{d}(t)^\top & \boldsymbol{e}(t)^\top \end{bmatrix}^\top$.
	Integrating from $0$ to $T$ shows $ $
	\begin{equation*}
		\begin{gathered}
			\boldsymbol{x}(T)^\top P \boldsymbol{x}(T) - \boldsymbol{x}(0)^\top P \boldsymbol{x}(0) + \lambda \int_0^T \begin{bmatrix} \boldsymbol{v}(t) \\ \boldsymbol{w}(t) \end{bmatrix}^\top M \begin{bmatrix} \boldsymbol{v}(t) \\ \boldsymbol{w}(t) \end{bmatrix} dt \\
			\leq \int_0^T \begin{bmatrix} \boldsymbol{d}(t) \\ \boldsymbol{e}(t) \end{bmatrix}^\top X \begin{bmatrix} \boldsymbol{d}(t) \\ \boldsymbol{e}(t) \end{bmatrix} dt
		\end{gathered}
	\end{equation*}
	where the left-hand side can be lower-bounded by $\boldsymbol{x}(T)^\top P \boldsymbol{x}(T) - \boldsymbol{x}(0)^\top P \boldsymbol{x}(0)$ due to the signal pair $(\boldsymbol{v}, \boldsymbol{w})$ satisfying the static IQC defined by $M$, and thus the integral involving $M$ being nonnegative. This certifies dissipativity with the storage function $S(x) = x^\top P x$.
\end{pf}

\subsection{Dynamic IQCs} \label{subsec:dynamic-iqc}

We now extend the result of the previous subsection to a class of dynamic IQCs.
Assume $\Delta$ satisfies the IQC defined by $(\Psi, M)$ where
$\Psi$ can be decomposed into a filter $\Psi_1$ that operates on $\boldsymbol{v}$ and a filter $\Psi_2$ that operates on $\boldsymbol{w}$, and $M$ can be decomposed as:
\begin{equation} \label{eq:dynamic-iqc-assumptions}
	\Psi = \begin{bmatrix} \Psi_1 & 0 \\ 0 & \Psi_2 \end{bmatrix},
	\quad
	M = \begin{bmatrix} I & 0 \\ 0 & -I \end{bmatrix},
\end{equation}
and $\Psi$ is invertible, stable, and has a stable proper inverse.
This class itself can be used to describe, for example, uncertain LTI dynamics \citep{megretski_system_1997}.
Methods from \citep{wang_seiler_lpv_2016} can be applied to extend what follows to a broader class of dynamic IQCs.

\begin{figure}[tb]
	\centering
	\includegraphics[width=0.8\linewidth]{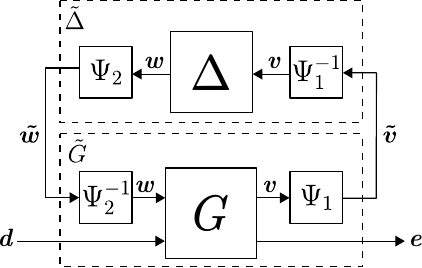}
	\caption{The original system, the interconnection of $G$ and $\Delta$, is transformed such that the transformed uncertainty $\tilde{\Delta}$ satisfies a simpler, static, IQC.}
	\label{fig:dynamic-iqc-transformation}
\end{figure}

Note that the operator $\tilde{\Delta} \triangleq \Psi_2 \Delta \Psi_1^{-1}$ satisfies the static IQC defined by $\begin{bsmallmatrix} I & 0 \\ 0 & -I \end{bsmallmatrix}$.
In the following, we manipulate \eqref{eq:uncertain-lti-sys} to be the interconnection of an LTI system with uncertainty $\tilde{\Delta}$, and then apply the methods from Section~\ref{subsec:static-iqc}.
See Figure~\ref{fig:dynamic-iqc-transformation} for a depiction of this transformation.

We write realizations of $\Psi_1$ and $\Psi_2$ as follows.
\begin{align}
	\begin{bmatrix} \dot{\boldsymbol{\psi}}_{\boldsymbol{1}}(t) \\ \boldsymbol{\tilde{v}}(t) \end{bmatrix} & = \begin{bmatrix} A_{\psi_1} & B_{\psi_1} \\ C_{\psi_1} & D_{\psi_1} \end{bmatrix} \begin{bmatrix} \boldsymbol{\psi_1}(t) \\ \boldsymbol{v}(t) \end{bmatrix}                             \\
	\begin{bmatrix} \dot{\boldsymbol{\psi}}_{\boldsymbol{2}}(t) \\ \boldsymbol{\tilde{w}}(t) \end{bmatrix} & = \begin{bmatrix} A_{\psi_2} & B_{\psi_2} \\ C_{\psi_2} & D_{\psi_2} \end{bmatrix} \begin{bmatrix} \boldsymbol{\psi_2}(t) \\ \boldsymbol{w}(t) \end{bmatrix} \label{eq:psi2-realization}
\end{align}

Defining $\tilde{x} \triangleq \begin{bmatrix}
		x^\top & \psi_1^\top & \psi_2^\top
	\end{bmatrix}^\top$, we form the extended system,
\begin{equation} \label{eq:dynamic-iqc-extended-system} 
	\begin{aligned}
		\begin{bmatrix} \dot{\tilde{\boldsymbol{x} } }(t) \\ \boldsymbol{v}(t) \\ \boldsymbol{\tilde{v}}(t) \\ \boldsymbol{e}(t) \end{bmatrix}
		                  & =
		\begin{bmatrix}
			\tilde{A}                                 & \tilde{B}_w     & \tilde{B}_d     \\
			\begin{bmatrix} C_v & 0 & 0 \end{bmatrix} & D_{v w}         & D_{v d}         \\
			\tilde{C}_v                               & \tilde{D}_{v w} & \tilde{D}_{v d} \\
			\tilde{C}_e                               & \tilde{D}_{e w} & \tilde{D}_{e d}
		\end{bmatrix} \begin{bmatrix} \boldsymbol{\tilde{x}}(t) \\ \boldsymbol{w}(t) \\ \boldsymbol{d}(t) \end{bmatrix} \\
		\boldsymbol{w}(t) & = \Delta(\boldsymbol{v})(t).
	\end{aligned}
\end{equation}
The expansions of $\tilde{A}, \dots, \tilde{D}_{e d}$ can be found in Appendix~\ref{appendix:tilde-expansion}.

Now, we swap $w$ with $\tilde{w}$ using $w = D_{\psi_2}^{-1}\tilde{w} -D_{\psi_2}^{-1} C_{\psi_2} \psi_2$ derived from equation~\eqref{eq:psi2-realization}, and eliminate the output $v$.

\begin{equation} 
	\begin{aligned}  \label{eq:dynamic-iqc-transformed-system}
		\begin{bmatrix} \dot{\tilde{\boldsymbol{x}}}(t) \\ \boldsymbol{\tilde{v}}(t) \\ \boldsymbol{e}(t) \end{bmatrix} & = \begin{bmatrix}
			                                                                                                                                   \tilde{A}   & \tilde{B}_w     & \tilde{B}_d     \\
			                                                                                                                                   \tilde{C}_v & \tilde{D}_{v w} & \tilde{D}_{v d} \\
			                                                                                                                                   \tilde{C}_e & \tilde{D}_{e w} & \tilde{D}_{e d}
		                                                                                                                                   \end{bmatrix}
		T
		\begin{bmatrix} \boldsymbol{\tilde{x}}(t) \\ \boldsymbol{\tilde{w}}(t) \\ \boldsymbol{d}(t) \end{bmatrix}                                              \\
		\boldsymbol{\tilde{w}}(t)                                                                                                      & = \tilde{\Delta}(\boldsymbol{\tilde{v}})(t)
	\end{aligned}
\end{equation}
where $T$ is defined as follows.
\begin{equation*}
	T \triangleq \begin{bmatrix}
		I                                                                 & 0               & 0 \\
		\begin{bmatrix} 0 & 0 & - D_{\psi_2}^{-1} C_{\psi_2}\end{bmatrix} & D_{\psi_2}^{-1} & 0 \\
		0                                                                 & 0               & I
	\end{bmatrix}
\end{equation*}

The transformed system in \eqref{eq:dynamic-iqc-transformed-system} is an uncertain LTI system where the uncertainty $\tilde{\Delta}$ satisfies a static IQC.
Further, it is well-posed since \eqref{eq:uncertain-lti-sys} is well-posed and $\Psi$ is both stable and has a stable proper inverse.
Thus, Lemma~\ref{lemma:static-iqc-dissipativity} provides a dissipativity condition for the transformed system. We now relate dissipativity of the extended and transformed system satisfying a static IQC to dissipativity of the original system satisfying a dynamic IQC.
\begin{lem} \label{lemma:dynamic-iqc-dissipativity}
	Suppose there exist $P \succeq 0$ and $\lambda \geq 0$ that satisfy the condition in Lemma~\ref{lemma:static-iqc-dissipativity} for the extended and transformed system in \eqref{eq:dynamic-iqc-transformed-system}, with $\tilde{\Delta}$ satisfying the static IQC defined by $\mathrm{diag}(I, -I)$.
	Then, the system defined in \eqref{eq:uncertain-lti-sys}, with the uncertainty satisfying a dynamic IQC of the form \eqref{eq:dynamic-iqc-assumptions}, is dissipative with respect to the quadratic supply rate in \eqref{eq:supply-rate}.
\end{lem}
\begin{pf}
	Suppose there exist $P \succeq 0$ and $\lambda \geq 0$ such that Lemma~\ref{lemma:static-iqc-dissipativity} is satisfied for the extended and transformed system in \eqref{eq:dynamic-iqc-transformed-system}.
	Then, there exists a storage function $S$ which certifies dissipativity of the extended and transformed system.
	This storage function also certifies dissipativity of the extended system in \eqref{eq:dynamic-iqc-extended-system} since an initial condition $\tilde{x}_0$ and disturbance signal  $\boldsymbol{d}$ induce the same trajectory $(\boldsymbol{\tilde{x}}, \boldsymbol{d}, \boldsymbol{e})$ in both systems.
	Thus, $S$ certifies dissipativity of the original system \eqref{eq:uncertain-lti-sys} with state consisting of the plant state and the virtual filter state.
\end{pf}

\section{Controller Synthesis} \label{sec:controller-synthesis}

In this section we present a procedure to synthesize a neural network controller that guarantees closed-loop dissipativity when in feedback with the plant design model in \eqref{eq:plant}, and maximizes a reward in the training environment.
First, we derive a matrix inequality that is equivalent to that used in Lemma~\ref{lemma:static-iqc-dissipativity} but is affine in variables corresponding to the controller variables, storage function, and controller activation function IQC multiplier.
This matrix inequality uses the plant design model.
Second, we present a constrained reinforcement learning procedure to train the controller to maximize reward, while using the dissipativity LMI to ensure dissipativity.
The environment used for training need not be the same as the plant design model; it can be a higher fidelity simulation or even the real system.

This controller synthesis problem can be posed as finding the controller $K$ that solves the optimization problem,
\begin{align*}
	K^* = \arg\max_K & \quad \mathbb{E}\left[ \int_0^T r(x(t), u(t)) dt \right] \\
	\mathrm{s.t.}    & \quad K\ \text{makes closed-loop dissipative }           \\
	\hphantom{s.t.}  & \quad K\ \text{is of the form in \eqref{eq:controller}},
\end{align*}
where the expectation is taken over a distribution of disturbance signals and of initial plant states.
This is the typical statement of a reinforcement learning problem \citep{suttonReinforcementLearningIntroduction2020} with the addition of the dissipativity constraint.
The expected reward is maximized over disturbances and initial plant states seen during training, subject to the controller satisfying a dissipativity constraint.

Following Lemma~\ref{lemma:dynamic-iqc-dissipativity}, we assume the plant design model has been extended and transformed to the form of \eqref{eq:plant} where $\Delta_p$ satisfies a static IQC.
We form the closed-loop system of the plant and controller:\
\begin{equation} \label{eq:closed-loop} 
	\begin{aligned}
		\begin{bmatrix} \dot{\boldsymbol{x}}(t) \\ \boldsymbol{v}(t) \\ \boldsymbol{e}(t) \end{bmatrix}
		                  & = \begin{bmatrix}
			                      A   & B_w     & B_d     \\
			                      C_v & D_{v w} & D_{v d} \\
			                      C_e & D_{e w} & D_{e d}
		                      \end{bmatrix}
		\begin{bmatrix} \boldsymbol{x}(t) \\ \boldsymbol{w}(t) \\ \boldsymbol{d}(t) \end{bmatrix} \\
		\boldsymbol{w}(t) & = \Delta(\boldsymbol{v})(t),
	\end{aligned}
\end{equation}
where $x = \begin{bmatrix} x_p^\top & x_k^\top \end{bmatrix}^\top$, $v = \begin{bmatrix} v_p^\top & v_k^\top \end{bmatrix}^\top$, $w = \begin{bmatrix} w_p^\top & w_k^\top \end{bmatrix}^\top$, and $\Delta: \boldsymbol{v} \mapsto \begin{bmatrix} \Delta_p(\boldsymbol{v_p})^\top & \phi(\boldsymbol{v_k})^\top \end{bmatrix}^\top$.
The matrices $A, \dots, D_{e d}$ are expressed in terms of the plant and controller matrices, and are affine in the controller matrices.
The expansions of these variables can be found in Appendix~\ref{appendix:closed-loop-expansion}.
The combined uncertainty $\Delta$ satisfies the static IQC defined by $M$ defined as
\begin{equation}\label{eq:M-decomp}
	M = \begin{bmatrix} M_{v v} & M_{v w} \\ M_{v w}^\top & M_{w w} \end{bmatrix} = \left[\begin{array}{cc|cc}
			M_{\Delta_p vv}       & 0       & M_{\Delta_p v w} & 0         \\
			0                     & 0       & 0                & \Lambda   \\ \hline
			M_{\Delta_p v w}^\top & 0       & M_{\Delta_p w w} & 0         \\
			0                     & \Lambda & 0                & -2\Lambda
		\end{array}\right].
\end{equation}

This closed-loop system is an uncertain LTI system of the form in \eqref{eq:uncertain-lti-sys}, with $\Delta$ satisfying a static IQC. Thus, we apply Lemma~\ref{lemma:static-iqc-dissipativity} to get a matrix inequality condition for dissipativity in terms of the closed-loop parameters and $M$.
For computational tractability of controller synthesis, we would like the matrix inequality to be linear in the controller matrices and storage function parameter $P$.
However, this condition is not affine in the controller matrices due to the term involving $M$ being quadratic in controller matrices when $M_{v v}$ is non-zero, and the term involving $X$ being quadratic when $X_{e e}$ is non-zero.

In the following, we let $\theta$ denote the controller parameters:
\begin{equation}\label{eq:RINN-parameters}
	\theta \triangleq \begin{bmatrix}
		A_k     & B_{k w}   & B_{k y}   \\
		C_{k v} & D_{k v w} & D_{k v y} \\
		C_{k u} & D_{k u w} & D_{k u y}
	\end{bmatrix}.
\end{equation}

The following subsections derive a bilinear matrix inequality (BMI) and then an LMI condition for dissipativity, and present a projection-based training algorithm for maximizing reward subject to the dissipativity condition.

\subsection{Bilinear Matrix Inequality (BMI) Condition for Dissipativity}

To convert the condition Lemma~\ref{lemma:static-iqc-dissipativity} into a BMI, we make two assumptions: that $X_{e e} \preceq 0$ and that $M_{\Delta_p v v} \succeq 0$.
This allows definition of $L_X$ and $L_{\Delta_p}$ such that
\begin{equation} \label{eq:semidefinite-assumptions}
	X_{e e} = -L_X^\top L_X, \quad M_{\Delta_p v v} = L_{\Delta_p}^\top L_{\Delta_p}.
\end{equation}
Then, we define $L_\Delta = \mathrm{diag}(L_{\Delta_p}, 0)$, such that $M_{v v} = L_\Delta^\top L_\Delta$.
Common quadratic supply rates that satisfy this negative semidefiniteness condition on $X_{e e}$ include those for $L_2$ gain and passivity.
Common IQC multipliers  that satisfy this positive semidefiniteness condition on $M_{\Delta_p v v}$ include those that characterize LTI dynamics, multiplication by a (possibly time-varying) real scalar, and memoryless nonlinearities sector-bounded in $[0, 1]$; see \citep{megretski_system_1997}.

Going forward, we let $\lambda = 1$.
This is without loss of generality since $\lambda$ can be absorbed into the $M_{\Delta_p}$ and $\Lambda$ variables.

With the semidefiniteness assumptions in \eqref{eq:semidefinite-assumptions}, we can apply the Schur complement to \eqref{eq:open-loop-dissipativity} to arrive at the  equivalent condition
\begin{equation}\label{eq:bmi} 
	\begin{bmatrix}
		F & \begin{bsmallmatrix}
			    C_v^\top  L_\Delta^\top & C_e^\top L_X^\top \\
			    D_{v w}^\top L_\Delta^\top & D_{e w}^\top L_X^\top \\
			    D_{v d}^\top  L_\Delta^\top & D_{e d}^\top L_X^\top
		    \end{bsmallmatrix} \\
		\begin{bsmallmatrix}
			L_\Delta C_v & L_\Delta D_{v w} & L_\Delta D_{v d} \\
			L_X C_e & L_X D_{e w} & L_X D_{e d}
		\end{bsmallmatrix}
		  & -I
	\end{bmatrix}
	\preceq 0
\end{equation}
where
\begin{align*}
	F = & \begin{bmatrix}
		      A^\top P + P A & P B_w & P B_d \\
		      B_w^\top P     & 0     & 0     \\
		      B_d^\top P     & 0     & 0
	      \end{bmatrix}                                                     \\
	    & + (\star)^\top \begin{bmatrix} 0 & M_{v w} \\ M_{v w}^\top & M_{w w}\end{bmatrix}
	\begin{bmatrix}
		C_v & D_{v w} & D_{v d} \\
		0   & I       & 0
	\end{bmatrix}                                                                  \\
	    & - (\star)^\top \begin{bmatrix} X_{d d} & X_{d e} \\ X_{d e}^\top & 0 \end{bmatrix}
	\begin{bmatrix}
		0   & 0       & I       \\
		C_e & D_{e w} & D_{e d}
	\end{bmatrix}.
\end{align*}

Since the matrices of the closed-loop system are affine in the controller parameters $\theta$, this condition is a bilinear matrix inequality in $P$, $\theta$, $\Lambda$, $L_{\Delta_p}$, $M_{\Delta_p v w}$, $M_{\Delta_p w w}$, $X_{d d}$, $X_{d e}$, and $L_X$.

\subsection{Partial Convexification of BMI}

We now use a change of variables to arrive at a matrix inequality that is linear in the new decision variables, such that $P$, $\Lambda$, and $\theta$ can be recovered from the new decision variables.
This change of variables is based on the one introduced in \citep{scherer_multiobjective_1997} for  $\mathcal{H}_\infty$ control.
For simplicity, we restrict the dimension of the controller state to match the dimension of the plant state, so $n_k = n_p$.
See \citep{scherer_multiobjective_1997} for notes on when the controller dimension differs from the plant dimension.

To apply this change of variables, we first require positive definiteness of the storage function parameter: $P \succ 0$.
Then, we partition $P$ and $P^{-1}$ as follows:
\begin{align*}
	P = \begin{bmatrix}
		    S & U \\ U^\top & \star
	    \end{bmatrix}
	\quad
	P^{-1} = \begin{bmatrix}
		         R & V \\ V^\top & \star
	         \end{bmatrix},
\end{align*}
where $S, R, U$, and $V \in \mathbb{R}^{n_p \times n_p}$.
Note that  $S \succ 0$ and $R \succ 0$.
We do not use the lower-right blocks of $P$ and $P^{-1}$ by name, so we label them with stars above.
The relevant properties of these blocks are that they are compatible with $S, R, U$, and $V$ such that $P \succ 0$ and $PP^{-1} = I$.
We further define
\begin{equation} \label{eq:y-def}
	Y \triangleq \begin{bmatrix} R & I \\ V^\top & 0 \end{bmatrix}.
\end{equation}
We now left- and right-multiply the matrix in \eqref{eq:bmi} by $\mathrm{diag}(Y^\top, I)$ and its transpose to get the condition
\begin{equation}\label{eq:yt-bmi-y} 
	\begin{bmatrix}
		\begin{bmatrix} Y^\top & 0 \\ 0 & I \end{bmatrix} F \begin{bmatrix} Y & 0 \\ 0 & I \end{bmatrix} &
		\begin{bsmallmatrix}
			Y^\top C_v^\top  L_\Delta^\top & Y^\top C_e^\top L_X^\top \\
			D_{v w}^\top L_\Delta^\top & D_{e w}^\top L_X^\top \\
			D_{v d}^\top  L_\Delta^\top & D_{e d}^\top L_X^\top
		\end{bsmallmatrix}                                             \\
		\begin{bsmallmatrix}
			L_\Delta C_v Y & L_\Delta D_{v w} & L_\Delta D_{v d} \\
			L_X C_e Y & L_X D_{e w} & L_X D_{e d}
		\end{bsmallmatrix}
		                                                                                                 & -I
	\end{bmatrix}
	\preceq 0
\end{equation}
where
\begin{gather*}
	\begin{bmatrix} Y^\top & 0 \\ 0 & I \end{bmatrix} F \begin{bmatrix} Y & 0 \\ 0 & I \end{bmatrix} = \\
	\begin{bmatrix}
		Y^\top A^\top P Y + Y^\top P A Y & Y^\top P B_w & Y^\top P B_d \\
		B_w^\top P Y                     & 0            & 0            \\
		B_d^\top P Y                     & 0            & 0
	\end{bmatrix} \\
	+ (\star)^\top \begin{bmatrix} 0 & M_{v w} \\ M_{v w}^\top & M_{w w}\end{bmatrix}
	\begin{bmatrix}
		C_v Y & D_{v w} & D_{v d} \\
		0     & I       & 0
	\end{bmatrix} \\
	- (\star)^\top \begin{bmatrix} X_{d d} & X_{d e} \\ X_{d e}^\top & 0 \end{bmatrix}
	\begin{bmatrix}
		0     & 0       & I       \\
		C_e Y & D_{e w} & D_{e d}
	\end{bmatrix}.
\end{gather*}

We construct the following variables from $P$, $\theta$, and $\Lambda$:
\begin{equation} \label{eq:change-of-vars}
	\begin{aligned}
		N_A             & \triangleq
		{\begin{bmatrix}
				 N_{A11} & N_{A12} \\ N_{A21} & N_{A22}
			 \end{bmatrix}}                                            \\ = &
		\begin{bmatrix}
			S A_p R & 0 \\ 0 & 0
		\end{bmatrix}
		+ \begin{bmatrix}
			  U & S B_{p u} \\ 0 & I
		  \end{bmatrix}
		\begin{bmatrix}
			A_k & B_{k y} \\ C_{k u} & D_{k u y}
		\end{bmatrix}
		\begin{bmatrix}
			V^\top & 0 \\ C_{p y} R & I
		\end{bmatrix}                                                        \\
		N_B             & \triangleq S B_{p u} D_{k u w} + U B_{k w}                      \\
		N_C             & \triangleq \Lambda D_{k v y} C_{p y} R + \Lambda C_{k v} V^\top \\
		\hat{D}_{k v y} & \triangleq \Lambda D_{k v y}                                    \\
		\hat{D}_{k v w} & \triangleq \Lambda D_{k v w}.
	\end{aligned}
\end{equation}
With these definitions, the new decision variables are $\hat{\theta} = \{S, R, N_A, N_B, N_C, D_{k u w}, \hat{D}_{k v y}, \hat{D}_{k v w}, \Lambda\}$.
The matrix in \eqref{eq:yt-bmi-y} is affine in $\hat{\theta}$, $M_{\Delta_p w w}$, and $X_{d d}$.
See  \eqref{eq:expansions} for the expansions of the terms in \eqref{eq:yt-bmi-y} in terms of the new decision variables.

\begin{table*}[btp]
	\begin{equation} \label{eq:expansions}
		\begin{gathered}
			\begin{gathered}
				Y^\top P A Y  = \begin{bmatrix}
					A_p R + B_{p u} N_{A 2 1} & A_p + B_{p u} N_{A 2 2} C_{p y} \\
					N_{A 1 1}                 & S A_p + N_{A 1 2} C_{p y}
				\end{bmatrix} \quad
				Y^\top P B_{w}  = \begin{bmatrix}
					B_{p w} + B_{p u} N_{A 2 2} D_{p y w} & B_{p u} D_{k u w} \\
					S B_{p w} + N_{A 1 2} D_{p y w}       & N_B
				\end{bmatrix}
			\end{gathered} \\
			\begin{gathered}
				Y^\top P B_d  = \begin{bmatrix}
					B_{p d} + B_{p u} N_{A 2 2} D_{p y d} \\
					S B_{p d } + N_{A 1 2} D_{p y d}
				\end{bmatrix}
				\quad
				M_{v w}^\top C_v Y  = \begin{bmatrix}
					M_{\Delta_p v w}^\top (C_{p v} R + D_{p v u} N_{A 2 1}) & M_{\Delta_p v w}^\top( C_{p v} +  D_{p v u} N_{A 2 2} C_{p y}) \\
					N_C                                                     & \hat{D}_{k v y} C_{p y}
				\end{bmatrix}\\
				- X_{d e} C_e Y   = - X_{d e} \begin{bmatrix}
					C_{p e} R + D_{p e u} N_{A 2 1} & C_{p e} + D_{p e u} N_{A 2 2} C_{p y}
				\end{bmatrix}
			\end{gathered} \\
			\begin{alignedat}{2}
				L_\Delta C_v Y       & = \begin{bsmallmatrix}
					                         L_{\Delta_p} \begin{bsmallmatrix}
						C_{p v} R + D_{p v u} N_{A 2 1} & C_{p v} + D_{p v u} N_{A 2 2} C_{p y}
					\end{bsmallmatrix} \\
					                         0
				                         \end{bsmallmatrix}                             & \quad
				L_X C_e Y            & = L_X \begin{bsmallmatrix}
					                             C_{p e} R + D_{p e u} N_{A 2 1} & C_{p e} + D_{p e u} N_{A 2 2} C_{p y}
				                             \end{bsmallmatrix} \\
				M_{v w}^\top D_{v w} & = \begin{bsmallmatrix}
					                         M_{\Delta_p v w}^\top (D_{p v w} + D_{p v u} N_{A 2 2} D_{p y w})
					                         & M_{\Delta_p v w}^\top D_{p v u} D_{k u w} \\
					                         \hat{D}_{k v y} D_{p y w} & \hat{D}_{k v w}
				                         \end{bsmallmatrix}       & \quad
				M_{v w}^\top D_{v d} & = \begin{bsmallmatrix}
					                         M_{\Delta_p v w}^\top (D_{p v d} + D_{p v u} N_{A 2 2} D_{p y d}) \\
					                         \hat{D}_{k v y} D_{p y d}
				                         \end{bsmallmatrix}
			\end{alignedat}
		\end{gathered}
	\end{equation}
\end{table*}

\begin{thm} \label{thm:controller-synthesis}
	For a fixed $M_{\Delta_p}$ and $X$, there exists a controller of the same order as the plant and a quadratic storage function that satisfies Lemma~\ref{lemma:static-iqc-dissipativity} with positive definite $P$ and $\Lambda$, certifying closed-loop dissipativity, if and only if there exist
	$\hat{\theta} = \{S, R, N_A, N_B, N_C, D_{k u w}, \hat{D}_{k v y}, \hat{D}_{k v w}, \Lambda\}$ that satisfy \eqref{eq:yt-bmi-y}
	where $S \succ 0$, $R \succ 0$, $\Lambda \succ 0$, and $\begin{bsmallmatrix}
			R & I \\ I & S
		\end{bsmallmatrix} \succ 0$.
\end{thm}
\begin{pf}
	First, we show necessity. Suppose there exists a controller and a $P \succ 0$ satisfying Lemma~\ref{lemma:static-iqc-dissipativity}, certifying closed-loop dissipativity.
	Then, \eqref{eq:bmi} holds by applying the Schur complement.
	This condition implies \eqref{eq:yt-bmi-y}.
	Defining variables as in \eqref{eq:change-of-vars} shows necessity.

	Now we show sufficiency. Suppose there exists a solution $\hat{\theta}$ to \eqref{eq:yt-bmi-y} where $S, R, \Lambda \succ 0$, and $\begin{bsmallmatrix} R & I \\ I & S \end{bsmallmatrix} \succ 0$. Note that $S \succ 0$ along with  $\begin{bsmallmatrix} R & I \\ I & S \end{bsmallmatrix} \succ 0$ implies $I - RS$ is invertible.
	First, construct $U$ and $V$ through SVD of $I - RS$ as follows: $V U^\top = I - RS$.
	Since $I - RS$ is invertible, so are $V$ and $U$.
	Since both $R$ and $V$ are invertible, so is $Y$.
	This gives a construction of $P$ by $\begin{bsmallmatrix} I & S \\ 0 & U^\top \end{bsmallmatrix} Y^{-1}$.
	Then, reconstruct $D_{k v y}$ and $D_{k v w}$ by solving for them in the equations for $\hat{D}_{k v y}$ and $\hat{D}_{k v w}$ in \eqref{eq:change-of-vars}.
	Next, the equations for $N_A$, $N_B$, and $N_C$ are solved for $(A_k, B_{k y}, C_{k u}, D_{k u y})$, $B_{k w}$, and $C_{k v}$ respectively.
	Due to the invertibility of $Y$, \eqref{eq:yt-bmi-y} is equivalent to \eqref{eq:bmi}.
	Finally, through the Schur complement, \eqref{eq:bmi} is equivalent to the original dissipativity condition provided by Lemma~\ref{lemma:static-iqc-dissipativity}. This completes the proof of sufficiency.
\end{pf}
Note that \eqref{eq:yt-bmi-y} is also affine in $M_{\Delta_p w w}$ and $X_{d d}$, so these may also be made decision variables.
For example, when expressing the supply rate for $L_2$ gain as $s(d, e) = \gamma \|d\|^2 - \|e\|^2$, this allows minimizing $\gamma$ subject to \eqref{eq:yt-bmi-y}.

\subsection{Training Procedure}\label{subsec:training}

Algorithm~\ref{alg:simple-training} gives a projection-based procedure for training a RINN controller to maximize the specified reward function, while satisfying the dissipativity condition.
It alternates between a reinforcement learning training step and a dissipativity-enforcing step.
This enables it to be easily integrated into reinforcement learning algorithms.

For brevity of notation, we let $\hat{\Theta}(M_{\Delta_p}, X)$ be the set of $\hat{\theta}$ that satisfy the conditions listed in Theorem~\ref{thm:controller-synthesis} given $M_{\Delta_p}$ and $X$, and $\Theta(M_{\Delta_p}, X, P, \Lambda)$ be the set of $\theta$ that satisfy the conditions listed in Lemma~\ref{lemma:static-iqc-dissipativity} (using the closed-loop matrices), given $M_{\Delta_p}$, $X$, and $\Lambda$.

The algorithm begins by initializing the controller, $P$, and $\Lambda$.
The controller may be randomly initialized or warm-started (e.g., with an LTI controller satisfying dissipativity or a neural network trained for reward).
The matrices $P$ and $\Lambda$ may be set to the identity, or, if the initial controller is certified dissipative, $P$ and $\Lambda$ can be initialized with the corresponding certificate.

In the training loop, the training step of the reinforcement learning algorithm changes the controller parameters, and the new controller may not satisfy the dissipativity condition.
Therefore, we include a dissipativity-enforcing step to modify the controller if necessary.
First, we check if there exist parameters $P$ and $\Lambda$ that certify the modified controller parameters satisfy the dissipativity condition (line~\ref{algline:dissipativity-check}).
If so, the controller needs no further modification.

If we cannot certify dissipativity, we apply a projection step to construct new controller variables which  satisfy the dissipativity condition (lines~\ref{algline:projection-start}-\ref{algline:theta-proj}).
First, we construct $\hat{\theta}^\prime$ from the current controller parameters and the most recent $P$ and $\Lambda$.
Then, in \textsc{ThetaHatProject} (line~\ref{algline:thetahatproject}), we construct a well-conditioned $\hat{\theta}$ that ensures dissipativity while remaining close to $\hat{\theta}^\prime$.
This involves first solving for the minimum projection error $\delta^* = \min_{\hat{\theta}} \|\hat{\theta} - \hat{\theta}^\prime \|_F$ s.t. $\hat{\theta} \in \hat{\Theta}(M_{\Delta_p}, X)$.
Then, to improve conditioning of the $\theta$ that is reconstructed from $\hat{\theta}$, we solve a projection problem again, modified to allow slight suboptimality in projection.
In particular, we solve for $\hat{\theta}$ and an auxiliary decision variable $\epsilon_{RS}$ to maximize $\epsilon_{RS}$ subject to $\hat{\theta} \in \hat{\Theta}(M_{\Delta_p}, X)$, $\begin{bsmallmatrix}
		R & I \\ I & S
	\end{bsmallmatrix} \succ \epsilon_{RS} I$, and $\|\hat{\theta} - \hat{\theta}^\prime \|_F^2 \leq \beta^2 {\delta^*}^2$.
Here, $\beta$ is a backoff factor that tunes suboptimality, e.g. $\beta = 1.1$ allows projection error to be at most 10\% greater than optimal.
The main purpose of this relaxed projection is to improve the conditioning of $I-RS$ and thus the conditioning of $P$, which is used in the following, final, step.
In this step, we construct $\theta$ from $\hat{\theta}$.
While one option would be to use the reconstruction process described in the proof of Theorem~\ref{thm:controller-synthesis}, we instead extract $P$ and $\Lambda$ from $\hat{\theta}$ and then project $\theta^\prime$ into the set of controllers certified to be dissipative by $P$ and $\Lambda$.
This ensures that the resulting $\theta$ is as close as possible to $\theta^\prime$.

The number of floating point operations used to solve the semidefinite programs in Algorithm~\ref{alg:simple-training} with interior-point methods scales with sizes of the system and controller as follows \citep{Boyd_Vandenberghe_2004}.
Assuming $n_p = n_k$ and $n_p, n_\phi \gg n_w, n_d, n_y, n_u, n_v, n_e$, the computational complexity of the dissipativity check is $\mathcal{O}((n_p^2 + n_\phi)^2(n_p + n_\phi)^2)$, of the $\hat{\theta}$-space projection is $\mathcal{O}((n_p + n_\phi)^6)$, and of the $\theta$-space projection is $\mathcal{O}((n_p + n_\phi)^6)$.
Note that the dissipativity check has lower-order terms in $n_\phi$, the number of activation functions, compared to the other semidefinite programs, and the other semidefinite programs are run only if the dissipativity check fails.

\begin{algorithm}
	\caption{Neural Network Controller Training}
	\label{alg:simple-training}
	\begin{algorithmic}[1]
		\State $\theta \gets$ arbitrary
		\State $P, \Lambda \gets I$ \label{algline:init}
		\For{$i=1,\dots$}
		\Statex \hspace{1.5em}$\triangleright$ \eqparbox{Comment}{Training step} \hfill
		\State $\theta^\prime \gets$ reinforcement learning step on $\theta$
		\Statex
		\Statex \hspace{1.5em}$\triangleright$ \eqparbox{Comment}{Dissipativity-enforcing step} \hfill
		\If{$\exists P^\prime, \Lambda^\prime : \theta^\prime$ is dissipative}\label{algline:dissipativity-check}
		\State $\theta, P, \Lambda \gets \theta^\prime, P^\prime, \Lambda^\prime$
		\Else
		\State $\hat{\theta}^\prime \gets$ construct $\hat{\theta}$ from $\theta^\prime, P$, and $\Lambda$ \label{algline:projection-start}
		\State $\hat{\theta} \gets \Call{ThetaHatProject}{\hat{\theta}^\prime, \hat{\Theta}(M_{\Delta_p}, X)}$ \label{algline:thetahatproject}
		\State $P, \Lambda \gets$ extract $P$ and $\Lambda$ from $\hat{\theta}$
		\State $\theta \gets \arg\min_{\theta} \|\theta - \theta^\prime\| : \theta \in \Theta(M_{\Delta_p}, X, P, \Lambda)$ \label{algline:theta-proj}
		\EndIf
		\EndFor
	\end{algorithmic}
\end{algorithm}

\begin{rem}
	An alternate training procedure is to train directly on $\hat{\theta}$ by constructing $\theta$ from $\hat{\theta}$ through the process described in the proof of Theorem~\ref{thm:controller-synthesis}, as in \citep{junnarkar_synthesis_2022}.
	This leverages the reinforcement learning algorithm to take gradients through the reconstruction process of $\theta$ from $\hat{\theta}$, and directly improve $\hat{\theta}$.
	However, compared with training directly on $\hat{\theta}$, we find Algorithm~\ref{alg:simple-training} to have less variance across seeds and to be less susceptible to sudden drops in reward.
	We attribute this to Algorithm~\ref{alg:simple-training} projecting directly in $\theta$-space, whereas training on $\hat{\theta}$ projects in $\hat{\theta}$-space and then transforms to $\theta$.
\end{rem}

\begin{rem}
	Although Algorithm~\ref{alg:simple-training} and previous sections assume that the neural network controller is in the form of a RINN for convenience of analysis, this is not needed for training.
	Instead, during training, the neural network controller can be converted into and out of the form of a RINN as needed for the dissipativity checks and projection step.
	A description of modeling neural networks in the implicit form can be found in \citep{el_ghaoui_implicit_2021}.
	Imposing specific neural network architectures amounts to constraining certain elements of $D_{k v w}$ (or $\hat{D}_{k v w}$) to be $0$.
\end{rem}

\begin{rem}
	If directly training an implicit neural network, well-posedness of the controller must be ensured.
	A convenient condition to use here is $\Lambda D_{kvw} + D_{kvw}^\top \Lambda - 2\Lambda \prec 0$, where $\Lambda \succ 0$ and diagonal.
	This condition imposes well-posedness when the activation function $\phi$ is monotone and slope-restricted in $[0, 1]$, which holds for common neural network activation functions \citep{revayRecurrentEquilibriumNetworks2023}.
	This constraint parallels the $\Lambda D_{kvw} + D_{kvw}^\top \Lambda - 2\Lambda \preceq 0$ constraint that appears in the dissipativity condition, but with definiteness instead of semidefiniteness.
	When applying conditions in the $\hat{\theta}$ variables, this condition becomes $\hat{D}_{kvw} + \hat{D}_{kvw}^\top - 2\Lambda \prec 0$, which is affine in $\hat{\theta}$.
	These constraints can be added to  $\hat{\Theta}(M_{\Delta_p}, X)$  and $\Theta(M_{\Delta_p}, X, P, \Lambda)$ to ensure well-posedness of the controllers synthesized during training.
\end{rem}

\begin{rem}
	To reduce training time, the dissipativity-enforcing step may be skipped in some iterations.
	Note that a dissipativity-enforcing step must be taken before using the controller to ensure dissipativity.
	In the primary experiments in this paper, we apply the dissipativity-enforcing step at each iteration of training for simplicity.
	In Section~\ref{sec:ablation}, we test the impact of this projection frequency.
\end{rem}

\subsection{Training via Regularization} \label{subsec:training-regularization}
A second training procedure we consider augments the reinforcement learning algorithm with a regularization term that penalizes violation of the dissipativity condition in Lemma~\ref{lemma:static-iqc-dissipativity}.
This method involves no SDPs and thus offers the potential of better scaling, although with addition of the hyperparameter $\lambda_s$, to which performance is sensitive.
However, we find performance of this method to be inconsistent across examples, as demonstrated in Section~\ref{sec:experiments}.

Let $E(\theta, P, M)$ denote the matrix in the condition \eqref{eq:bmi}, rendering the dissipation inequality as \(E(\theta, P, M) \preceq 0\).
For this training procedure, we use proximal policy optimization (PPO)~\citep{schulman2017proximalpolicyoptimizationalgorithms} and add the following regularization term to the policy loss function, where \(\lambda_s \in \mathbb{R}_{\geq 0}\) is a weighting hyperparameter.
\begin{equation*}
	L(\theta, P, M) = \lambda_s \max\left\{0, \lambda_{\max}(E(\theta, P, M))\right\}
\end{equation*}
This loss term penalizes positive eigenvalues of $E(\theta, P, M)$, incentivizing $E(\theta, P, M) \preceq 0$ and thus the dissipativity condition.
The controller model is augmented with learnable parameters $P$ and $M$ to enable training of the certificates for the controller $\theta$.
The matrix $P$ is parameterized as $\tilde{P}^\top \tilde{P}$ where $\tilde{P}$ is free to ensure $P \succeq 0$.
The matrix $M$ is constructed as in \eqref{eq:M-decomp}, with sub-blocks parameterized as per their IQC classes.

\section{Simulation} \label{sec:experiments}

We demonstrate the performance of this neural network controller approach through simulation on an inverted pendulum and a flexible rod on a cart.
We compare the reward of our neural network controller with dissipativity guarantees against a full-order LTI controller with the same guarantees.
To examine whether any loss in performance comes from neural network architecture or from dissipativity constraints, we additionally test two unconstrained neural network methods: a fully connected neural network and a standard, unconstrained RINN. These models do not guarantee closed-loop properties.
We refer to a RINN trained through Algorithm~\ref{alg:simple-training} as a Dissipative RINN (D-RINN), an unconstrained RINN as a Standard RINN (S-RINN), the LTI controller as LTI, and the fully connected neural network as FCNN.

We define plant models, the RINN controllers, and the LTI controllers in continuous time, and implement them by discretizing with the Runge-Kutta 4th order method to simulate between time steps.
We simulate the controller and plant independently between time steps, holding all inputs constant over each interval.
An additional implementation consideration is tuning the condition number of $U$ and $V$ to avoid ill-conditioned inverses.
While the suboptimal projection mentioned in Section~\ref{subsec:training} largely fixes this problem, we further improve conditioning by modifying the condition $\begin{bsmallmatrix} R & I \\ I & S \end{bsmallmatrix} \succ 0$ to be
\begin{equation} \label{eq:trs-def}
	\begin{bmatrix} R & t_{RS}I \\ t_{RS}I & S \end{bmatrix} \succ 0,
\end{equation}
and selecting $t_{RS}$ experimentally \citep{scherer_multiobjective_1997}.

Although for these numerical experiments we use simulation for both training and evaluation, the primary purpose of simulation is to tune the parameters of the controller to improve reward.
Thus, we distinguish between a training simulation, used for tuning the parameters of the controller to improve behavior, and the evaluation simulation, which is intended to be a representation of the actual system.
A modification we have found useful for training, and use in Section~\ref{subsec:flex-arm}, is to include a saturation on the control input.
We do not implement such modifications in the evaluation environment.
Note that, due to discretization in simulation, closed-loop dissipativity is not guaranteed in simulation.

All models are implemented in PyTorch for use in the RLLib reinforcement learning framework.
We use RLLib's implementation of proximal policy optimization, modified to include a dissipativity-enforcing step, as specified in Algorithm~\ref{alg:simple-training}.
We train the RINN models without specifying any structure on $D_{k v w}$.
The resulting implicit layer is implemented using TorchDEQ, which uses iterative methods to find the fixed point of the implicit equations.
Both the RINN models use the tanh activation function for all scalar nonlinearities.
All code is available at \url{https://github.com/neelayjunnarkar/neural-network-dissipativity}.

\subsection{Nonlinear Inverted Pendulum} \label{subsec:inv-pend}

In this example we consider the following inverted pendulum model:
\begin{align*}
	\dot{\boldsymbol{x}}_{\boldsymbol{1}}(t) & = \boldsymbol{x_2}(t)                                                                                                         \\
	\dot{\boldsymbol{x}}_{\boldsymbol{2}}(t) & = -\frac{\mu}{m \ell^2} \boldsymbol{x_2}(t) + \frac{g}{\ell} \sin(\boldsymbol{x_1}(t)) + \frac{1}{m \ell^2} \boldsymbol{u}(t) \\
	\boldsymbol{y}(t)                        & = \boldsymbol{x_1}(t)
\end{align*}
where $x_1$ is the angle of the pendulum from vertical, $x_2$ is the angular velocity of the pendulum, $u$ is the control input torque, $m=0.15$ kg is the mass, $\ell = 0.5$ m is the length of the pendulum, and $\mu = 0.05$ Nms/rad is the coefficient of friction.
This is modeled in the form of \eqref{eq:plant} by setting $\Delta_p(x_1) = \sin(x_1)$.
We characterize this uncertainty by the sector bound of $[0, 1]$, which holds over $[-\pi, \pi]$, and construct $M_{\Delta_p} = \begin{bsmallmatrix}
		0 & 1 \\ 1 & -2
	\end{bsmallmatrix}$.


We compare the performances of the controllers on this inverted pendulum with a reward constructed to incentivize minimal control effort.
The reward for each step of a rollout, which is of length 201, is set to $r(x, u) = -u^2/200$.
Initial conditions for the inverted pendulum are sampled uniformly at random with $x_1 \in [-0.6\pi, 0.6\pi]$ rad and $x_2 \in [-2, 2]$ rad/s.
If, during a rollout, $|x_1| > \pi$ rad or $|x_2| > 8$ rad/s, then the rollout is terminated and the remainder of the rollout is given a reward of 0.
We simulate using a time step of $0.01$ seconds.

We train a Dissipative RINN as per Algorithm~\ref{alg:simple-training}, with $e = \begin{bsmallmatrix} x_1 \\ x_2 \end{bsmallmatrix}$ and $X = 0$.
This guarantees closed-loop stability.
The state size, $n_k$, of the Dissipative RINN is the same as that of the plant, and the activation function size, $n_\phi$, is 16.
We initialize the Dissipative RINN with an LTI controller.
The LTI controller is synthesized by projecting $\hat{\theta}$, constructed with $(\theta, P, \Lambda) = (0, I, 0)$, onto the dissipative set restricted to LTI controllers, then recovering $\theta$ as in the proof of Theorem~\ref{thm:controller-synthesis}.
We use a $t_{RS}$ value of 1.5 in \eqref{eq:trs-def} for all projections, and train this model using a learning rate of $10^{-3}$.

\begin{figure}[tbp]
	\centering
	\includegraphics[width=\linewidth]{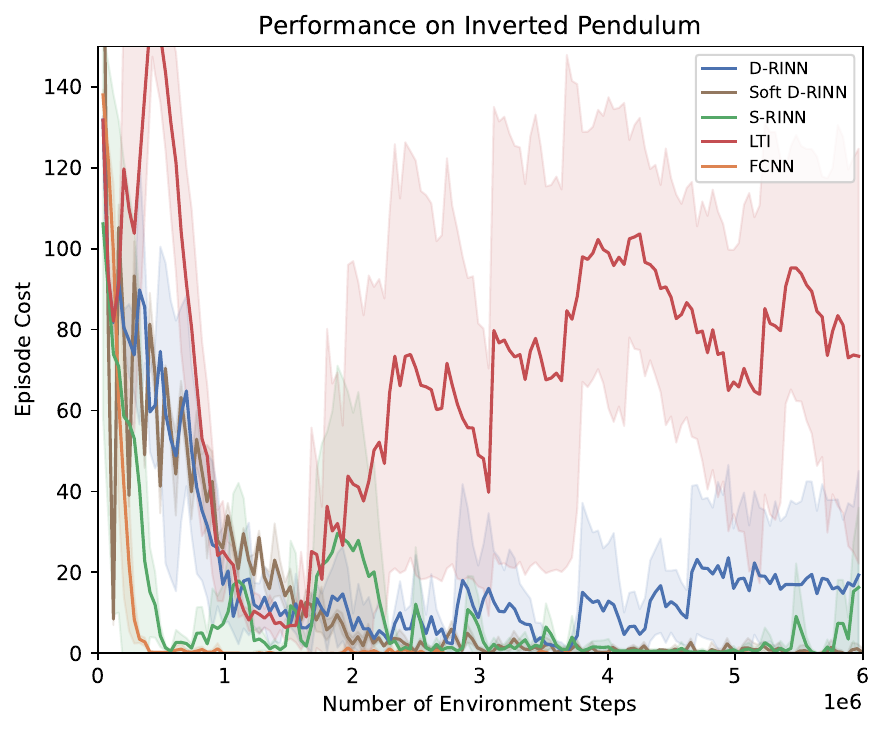}
	\caption{Episode cost vs. number of training environment steps for the Dissipative Recurrent Implicit Neural Network (D-RINN, our method), Soft D-RINN, Fully Connected Neural Network (FCNN), Standard RINN (S-RINN), and linear time-invariant (LTI) controllers on the inverted pendulum.
		The solid line represents the mean, and the shaded region represents 1 standard deviation.
		The cost is the sum of squared control inputs over the trajectory.
		The D-RINN and LTI controllers both guarantee stability of the inverted pendulum.
		The Soft D-RINN incentivizes but does not guarantee stability.
	}
	\label{fig:inverted-pendulum}
\end{figure}
We compare the Dissipative RINN with the following:
\begin{itemize}
	\item LTI: This model has the same state size as the Dissipative RINN.
	      It is initialized using the same method as used for the Dissipative RINN, and is trained using Algorithm~\ref{alg:simple-training} to satisfy the same constraints as the Dissipative RINN, with the only difference being that the $\hat{\theta}$ and $\theta$ projection steps are restricted so that they correspond only to LTI controllers.
	      We use the same $t_{RS}$ value in \eqref{eq:trs-def} as for the Dissipative RINN.
	      We use a learning rate of $10^{-2}$ for this model since it gives faster convergence without a performance penalty.

	\item Standard RINN: This model has the same state and activation function sizes as the Dissipative RINN.
	      After each training step, $\|D_{kvw}\|_\infty$ is reduced to be less than 1 to guarantee well-posedness  \citep{el_ghaoui_implicit_2021}. We use a learning rate of $10^{-3}$ for this model.

	\item Soft D-RINN: This model uses the regularization-based training described in Section~\ref{subsec:training-regularization}, with the same state and activation function sizes as the Dissipative RINN,
	      and regularization weight $\lambda_s = 0.1$.
	      We use a learning rate of $10^{-3}$ and initialize with the same LTI warm-start as the Dissipative RINN.

	\item Fully Connected Neural Network:
	      The fully connected neural network has two hidden layers, both with size 19, to give a total number of parameters roughly the same as the two RINN controllers.
	      We use a learning rate of $10^{-4}$ for this model.

\end{itemize}

\begin{figure*}[!t]
	\centering
	\begin{subfigure}[b]{0.3\textwidth} 
		\centering
		\includegraphics[width=\textwidth]{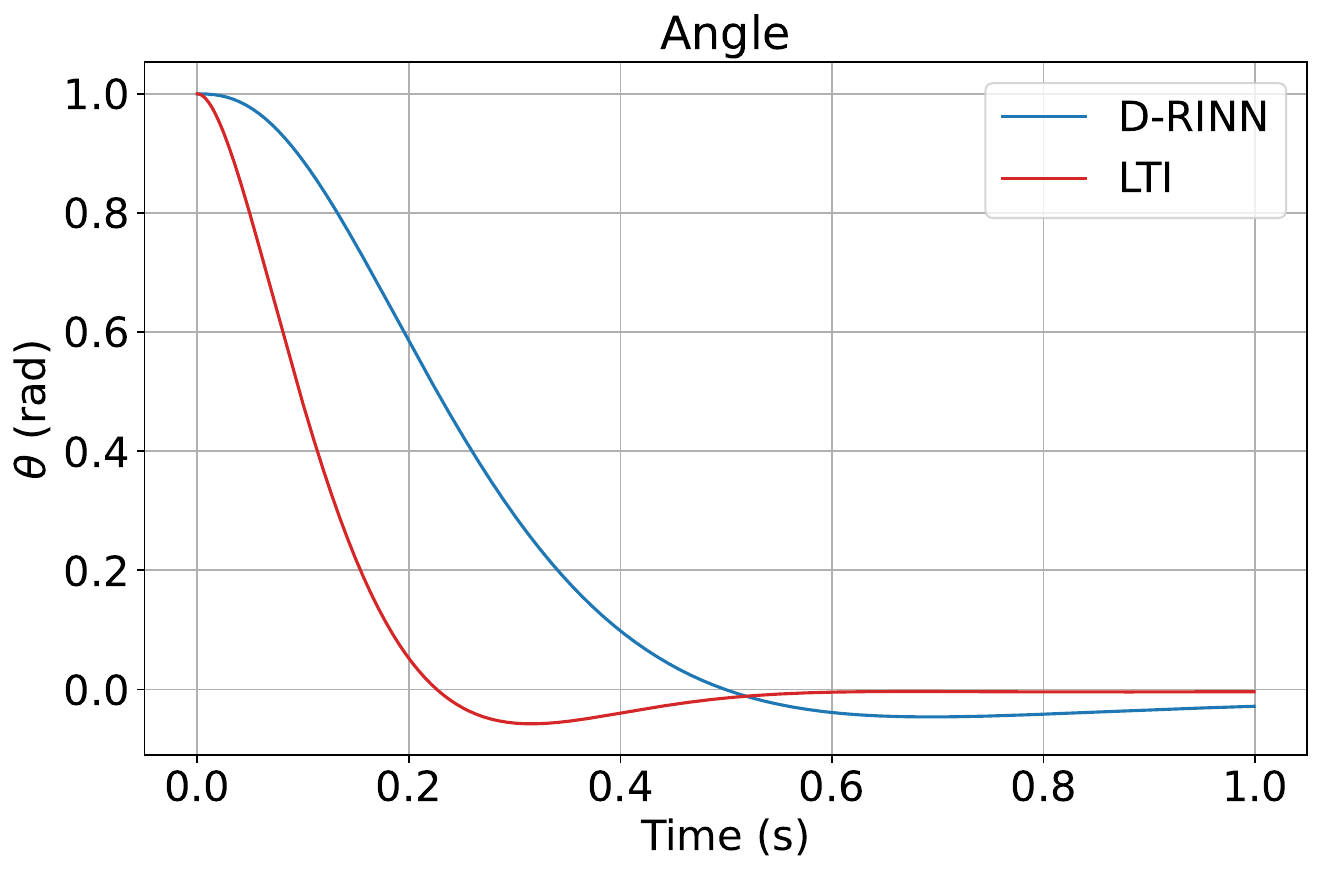}
		\label{subfig:inv-pend-time-simulation-angle}
	\end{subfigure}
	\hfill 
	\begin{subfigure}[b]{0.3\textwidth}
		\centering
		\includegraphics[width=\textwidth]{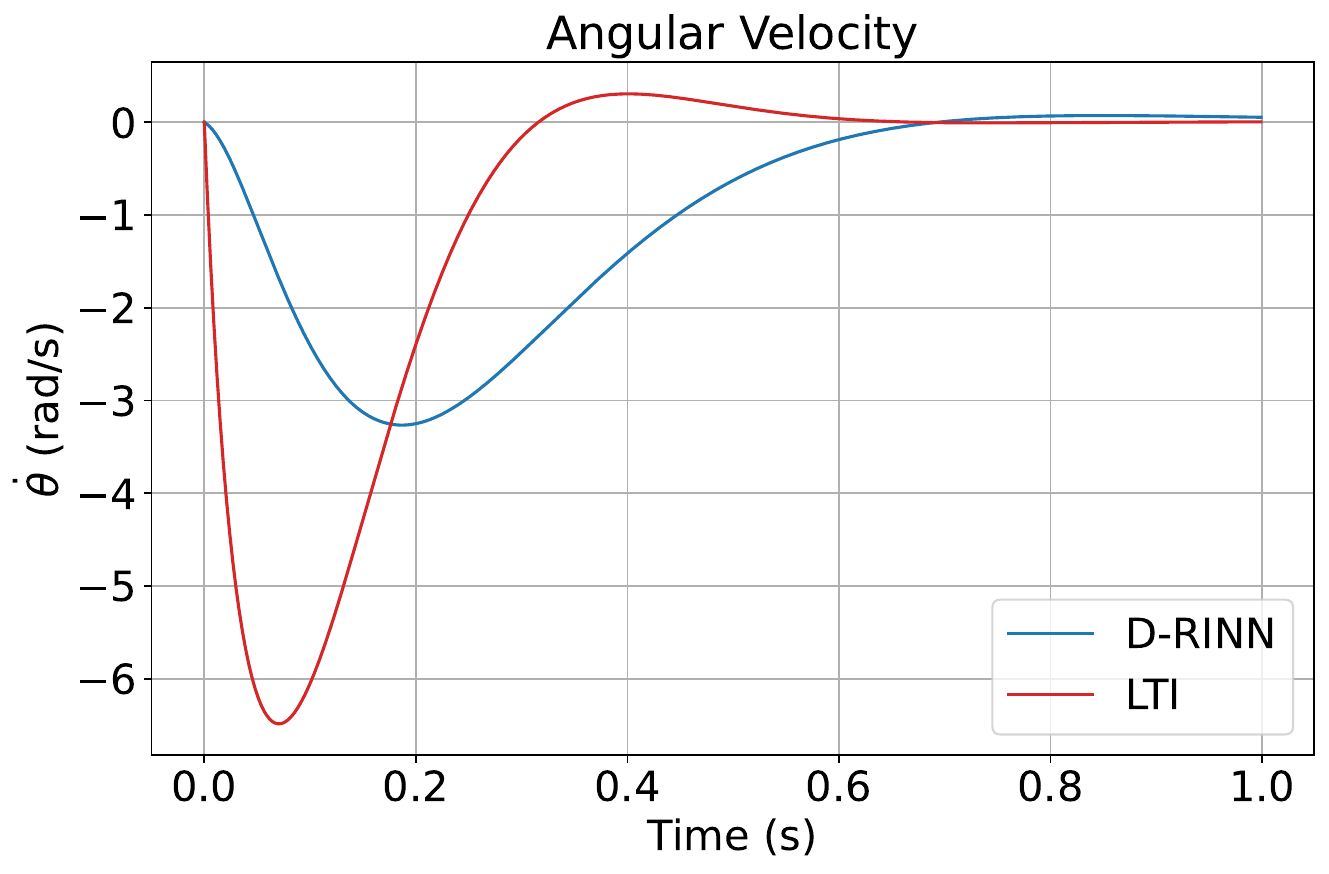}
		\label{subfig:inv-pend-time-simulation-velocity}
	\end{subfigure}
	\hfill
	\begin{subfigure}[b]{0.3\textwidth}
		\centering
		\includegraphics[width=\textwidth]{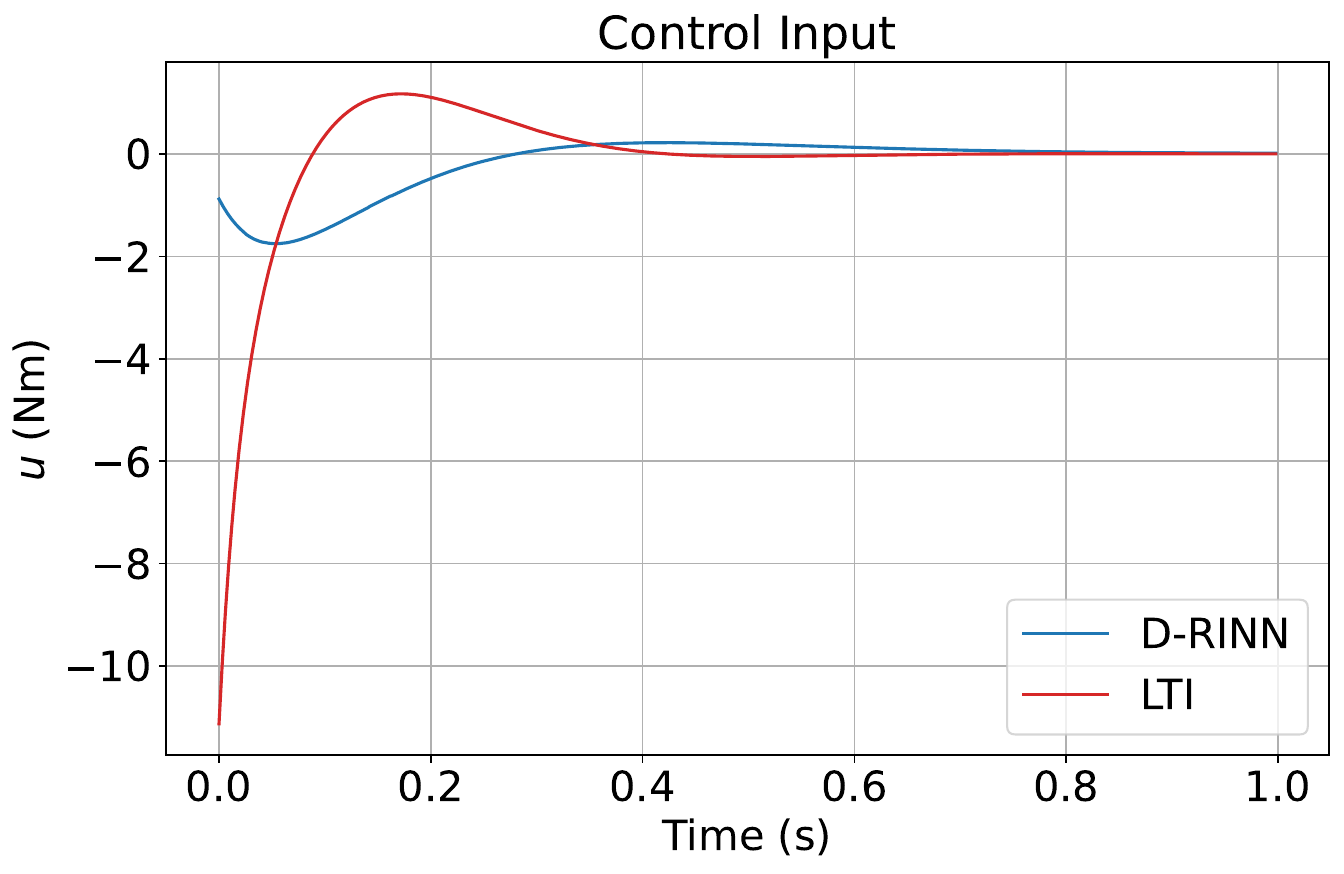}
		\label{subfig:inv-pend-time-simulation-action}
	\end{subfigure}
	\caption{
		Time simulations for the Dissipative Recurrent Implicit Neural Network (D-RINN, ours) and LTI controller starting from initial condition $\mathbf{x}(0)=\begin{bsmallmatrix}1 & 0\end{bsmallmatrix}^\top$ on the inverted pendulum.
		The D-RINN uses less control effort than the LTI controller to stabilize the system while satisfying the same stability guarantees.
	}
	\label{fig:inv-pend-time-simulation}
\end{figure*}

Each controller model is trained three times with different seeds for random number generation.
Figure~\ref{fig:inverted-pendulum} plots the training episode cost (negative of reward, summed over a trajectory) vs. the number of environment samples for the five controller models.
During training, the Standard RINN and fully connected neural network achieve near-zero cost quickly, as expected since the reward function incentivizes minimizing control effort, and these unconstrained controllers simply learn to output 0.
The Dissipative RINN and LTI controllers both guarantee closed-loop stability, and due to a common initialization method, start with similar reward.
Through training, the Dissipative RINN achieves episode cost a tenth of that achieved by the LTI controller.
The cost achieved by the Dissipative RINN is close to that achieved by its unconstrained counterpart, the Standard RINN.
The Soft D-RINN achieves episode cost comparable to the Dissipative RINN in early training, and exhibits better training stability in later stages.
Unlike the Dissipative RINN, the Soft D-RINN does not guarantee closed-loop stability, as the regularization term only incentivizes it.
We tested $\lambda_s \in \{0.01, 0.1, 1.0, 10.0\}$ and selected the smallest one for which the final controller satisfied the dissipativity condition.

Figure~\ref{fig:inv-pend-time-simulation} shows time simulations for the best trained controllers of the LTI and D-RINN models with the initial state of the inverted pendulum being $\mathbf{x_1}(0)=1$ and $\mathbf{x_2}(0) = 0$.
In the time simulations, we observe that the Dissipative RINN, our method, stabilizes the inverted pendulum with less control effort than the LTI controller, though it takes longer to stabilize.
This trade-off is expected because the reward function only penalizes control effort and not state size.
We do not include the unconstrained neural network controllers in the time simulation since they, as expected given the reward function, learn to output 0 and do not stabilize the system.

\subsubsection{Sensitivity to Hyperparameters}\label{sec:ablation}
We test the sensitivity of our proposed training method for the Dissipative RINN to the following hyperparameters: the number of activation functions $n_\phi$, the projection frequency, and the conditioning parameter $t_{RS}$.
For each configuration, we train the controller with three seeds and report the mean and standard deviation of the episode cost, averaged over the evaluations in the last 10\% of training time steps.
Results are reported in Table~\ref{tab:hyperparameter-sensitivity}, with values normalized in $[0, 1]$ relative to maximum possible cost, with lower cost being better.
For the projection frequency and $t_{RS}$, we see similar performance near the nominal values, but degradation in performance at large values.
For the neural network size, we see training instability at large values.

\begin{table}[htbp]
	\caption{D-RINN hyperparameter sensitivity on the inverted pendulum: mean and standard deviation of episode cost (normalized to $[0,1]$; lower is better) over three seeds.}
	\label{tab:hyperparameter-sensitivity}
	\centering
	\begin{tabular}{lrr}
		\hline
		Condition                 & Mean cost & Std    \\
		\hline
		\textit{NN size $n_\phi$} &           &        \\
		$n_\phi = 8$              & $0.00$    & $0.00$ \\
		$n_\phi = 16$ (nominal)   & $0.05$    & $0.04$ \\
		$n_\phi = 32$             & $0.27$    & $0.09$ \\
		\hline
		\textit{Proj.\ frequency} &           &        \\
		$1$ (nominal)             & $0.02$    & $0.02$ \\
		$5$                       & $0.05$    & $0.04$ \\
		$20$                      & $0.23$    & $0.15$ \\
		\hline
		\textit{$t_{RS}$}         &           &        \\
		$t_{RS} = 1.0$            & $0.05$    & $0.04$ \\
		$t_{RS} = 1.5$ (nominal)  & $0.03$    & $0.02$ \\
		$t_{RS} = 2.0$            & $0.24$    & $0.15$ \\
		\hline
	\end{tabular}
\end{table}

\subsection{Flexible Rod on a Cart} \label{subsec:flex-arm}

\begin{figure}[tpb]
	\centering
	\includegraphics[width=0.4\linewidth]{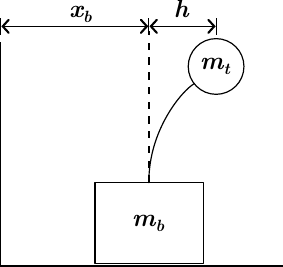}
	\caption{Diagram of the flexible rod on a cart.}
	\label{fig:flexarm-diagram}
\end{figure}

In this example, we consider a flexible rod on a cart (see Figure~\ref{fig:flexarm-diagram} for a diagram).
This consists of a movable cart with a metal rod fixed on top and a mass at the top of the rod.
Unlike the typical cart pole setup, there is no joint at the base of the rod.
However, the rod is flexible, so the horizontal position of the top and bottom of the rod can differ.
Training is conducted with the higher-fidelity flexible model \eqref{eq:flexarm-flexible-model} below, but $L_2$ gain constraints (where applied) are formulated with respect to a simplified model where the rod is rigid, allowing some uncertainty.
We compare the Dissipative RINN, LTI, Standard RINN, and fully connected neural network controllers on a reward function that rewards a combination of the state of the flexible rod on a cart being close to 0 and of the control effort being small.

The true dynamics, with the flexible rod, are taken as:
\begin{equation} \label{eq:flexarm-flexible-model}
	\begin{aligned}
		\dot{\boldsymbol{x}}_{\boldsymbol{f}}(t) & = \begin{bmatrix}
			                                             0 & I \\ -M_f^{-1} K_f & -M_f^{-1} B_f
		                                             \end{bmatrix} \boldsymbol{x_f}(t)
		+ \begin{bmatrix}
			  0 \\ M_f^{-1} \begin{bsmallmatrix}
				1 \\ 0
			\end{bsmallmatrix}
		  \end{bmatrix} \boldsymbol{u}(t)                                                                                                 \\
		\boldsymbol{y}(t)                        & = \begin{bmatrix}
			                                             1 & 1 & 0 & 0
		                                             \end{bmatrix} \boldsymbol{x_f}(t)                                                             \\
		M_f                                      & = \begin{bmatrix}
			                                             m_b + m_r + \rho L & m_t + \frac{\rho L}{3} \\ m_t + \frac{\rho L}{3} & m_t + \frac{\rho L}{5}
		                                             \end{bmatrix} \\
		K_f                                      & = \begin{bmatrix}
			                                             0 & 0 \\ 0 & \frac{4EI}{L^3}
		                                             \end{bmatrix}, \quad
		B_f                     = \begin{bmatrix}
			                          0 & 0 \\ 0 & 0.9
		                          \end{bmatrix}
	\end{aligned}
\end{equation}
where $x_f = \begin{bmatrix} x_b & h & \dot{x_b} & \dot{h} \end{bmatrix}^\top \in \mathbb{R}^4$ is the state of the flexible rod on a cart, $x_b$ is the position of the base of the rod, $h$ is the horizontal deviation of the top of the rod from the base of the rod, $m_b = 1$ kg is the mass of the base, $m_t = 0.1$ kg is the mass of an object at the top of the rod, $L = 1$ m is the length of the rod, $\rho = 0.1$ N/m is the mass density of the rod, $r = 10^{-2}$ m is the radius of the rod cross-section, $E = 200 \cdot 10^{-9}$ GPa is the Young's modulus of steel, and $I = \frac{\pi}{4} r^4$ m$^4$ is the area second moment of inertia.
The output $y$ represents the measurement of the position of the tip of the rod.

\begin{figure}[tbp]
	\centering
	\includegraphics[width=\linewidth]{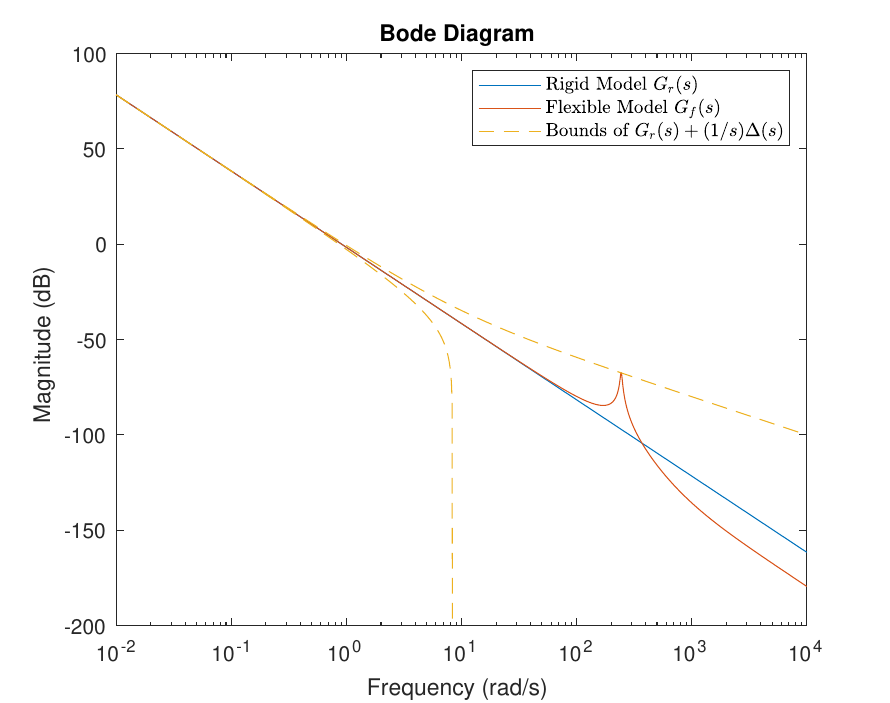}
	\caption{
		Bode magnitude plot, for the flexible rod on a cart, of the rigid model, the flexible model, and the bounds on the uncertain rigid model.
		The flexible model lies within the bounds of the uncertain rigid model.
	}
	\label{fig:flexarm-uncertainty-mag-bounds}
\end{figure}

The Dissipative RINN and LTI controllers are designed with respect to the following uncertain rigid model:
\begin{equation*}
	\begin{aligned}
		\dot{\boldsymbol{x}}_{\boldsymbol{r}}(t) & = \begin{bmatrix}
			                                             0 & 1 \\ 0 & 0
		                                             \end{bmatrix} \boldsymbol{x_r}(t)          \\
		                                         & \hphantom{=}
		+ \begin{bmatrix}
			  0 \\ \frac{1}{m_b + m_r + \rho L}
		  \end{bmatrix} ( \boldsymbol{u}(t) + \boldsymbol{d}(t))
		+ \begin{bmatrix} 1 \\ 0 \end{bmatrix} \boldsymbol{w}(t)
		\\
		\boldsymbol{w}(t)                        & = \Delta(\boldsymbol{u} + \boldsymbol{d})(t) \\
		\boldsymbol{y}(t)                        & = \boldsymbol{x_b}(t)                        \\
		\boldsymbol{e}(t)                        & = \boldsymbol{x_r}(t),
	\end{aligned}
\end{equation*}
where $x_r = \begin{bmatrix} x_b & \dot{x}_b\end{bmatrix}^\top$ and $\|\Delta\| \leq 0.1$.
The norm-bounded $\Delta$ block is used to represent uncertainty in the dynamical model.
If the rigid model with no uncertainty ($\Delta = 0$) has a transfer function from input $\boldsymbol{u+d}$ to output $\boldsymbol{y}$ of $G_r(s)$, then this uncertain model represents transfer functions of the form $G_r(s) + \frac{1}{s}\Delta(s)$ where $\|\Delta(s)\| \leq 0.1$.
The value of $0.1$ is chosen such that this set includes $G_f(s)$, the transfer function from input to output of the flexible model.
This ensures that a controller designed with closed-loop properties for this uncertain rigid model will satisfy the same properties for the flexible model.
See Figure~\ref{fig:flexarm-uncertainty-mag-bounds} for a Bode magnitude plot of $G_r(s)$, $G_f(s)$, and the bounds on $G_r(s) + \frac{1}{s} \Delta(s)$.

\begin{figure}[t]
	\centering
	\includegraphics[width=\linewidth]{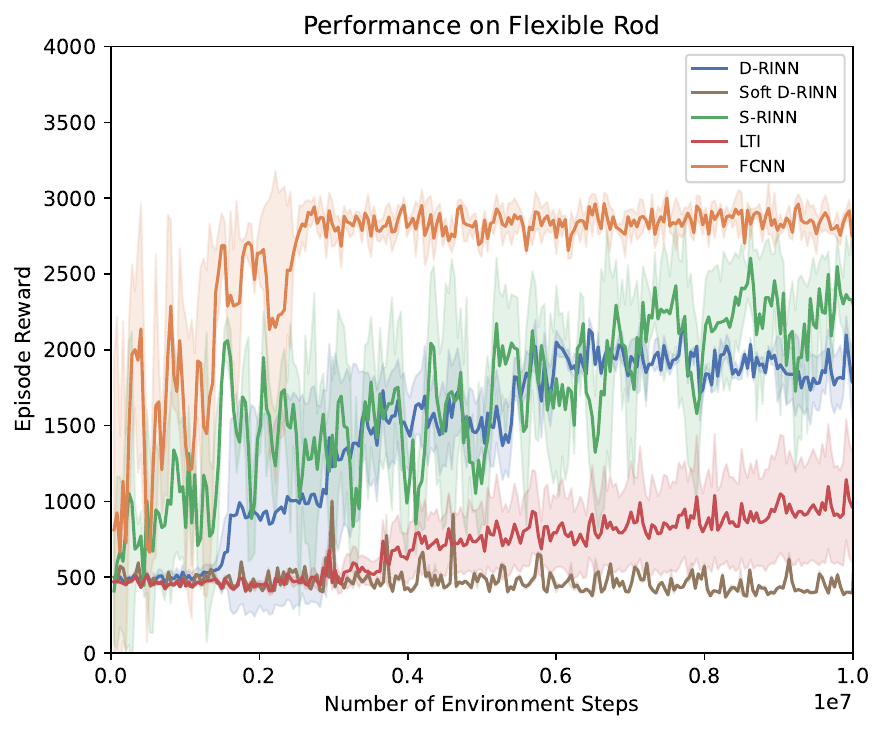}
	\caption{
		Evaluation reward vs. number of training environment steps for the Dissipative Recurrent Implicit Neural Network (D-RINN, our method), Soft D-RINN, Fully Connected Neural Network (FCNN), Standard RINN (S-RINN), and linear time-invariant (LTI) controllers on the flexible rod on a cart.
		The solid line represents the mean, and the shaded region represents 1 standard deviation.
		The reward is upper bounded by 4000, and lower bounded by 0.
		Higher reward indicates a combination of lower control effort and the state of the flexible rod on a cart being close to 0.
		The D-RINN and LTI controllers both guarantee an $L_2$ gain of at most 0.99 from a disturbance in the control input to the state for the uncertain rigid model.
		The Soft D-RINN incentivizes but does not guarantee this property.
	}
	\label{fig:flexarm}
\end{figure}

\begin{figure*}[!t]
	\centering
	\begin{subfigure}[b]{0.3\textwidth} 
		\centering
		\includegraphics[width=\textwidth]{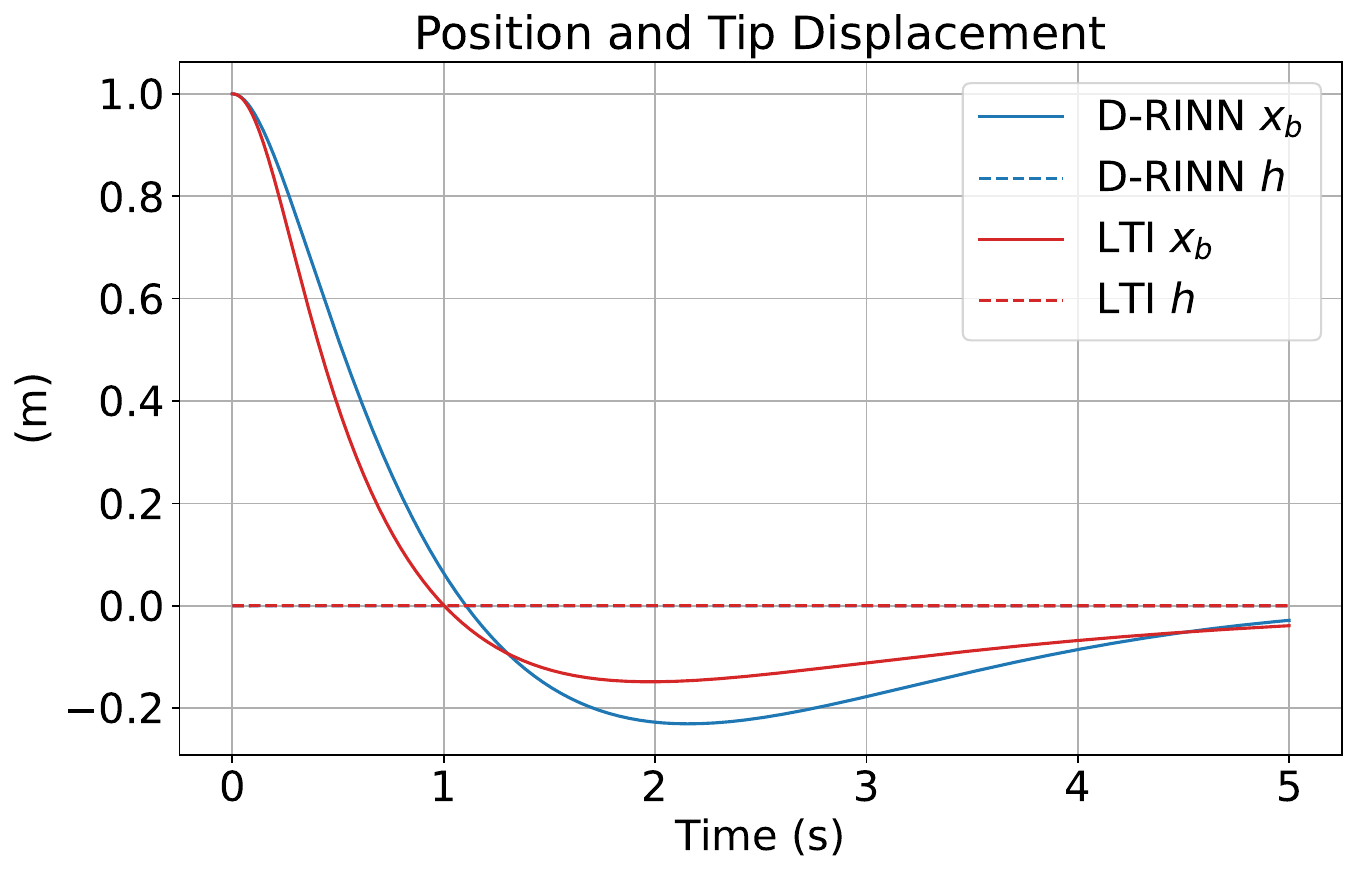}
		\label{subfig:flex-arm-time-simulation-position}
	\end{subfigure}
	\hfill 
	\begin{subfigure}[b]{0.3\textwidth}
		\centering
		\includegraphics[width=\textwidth]{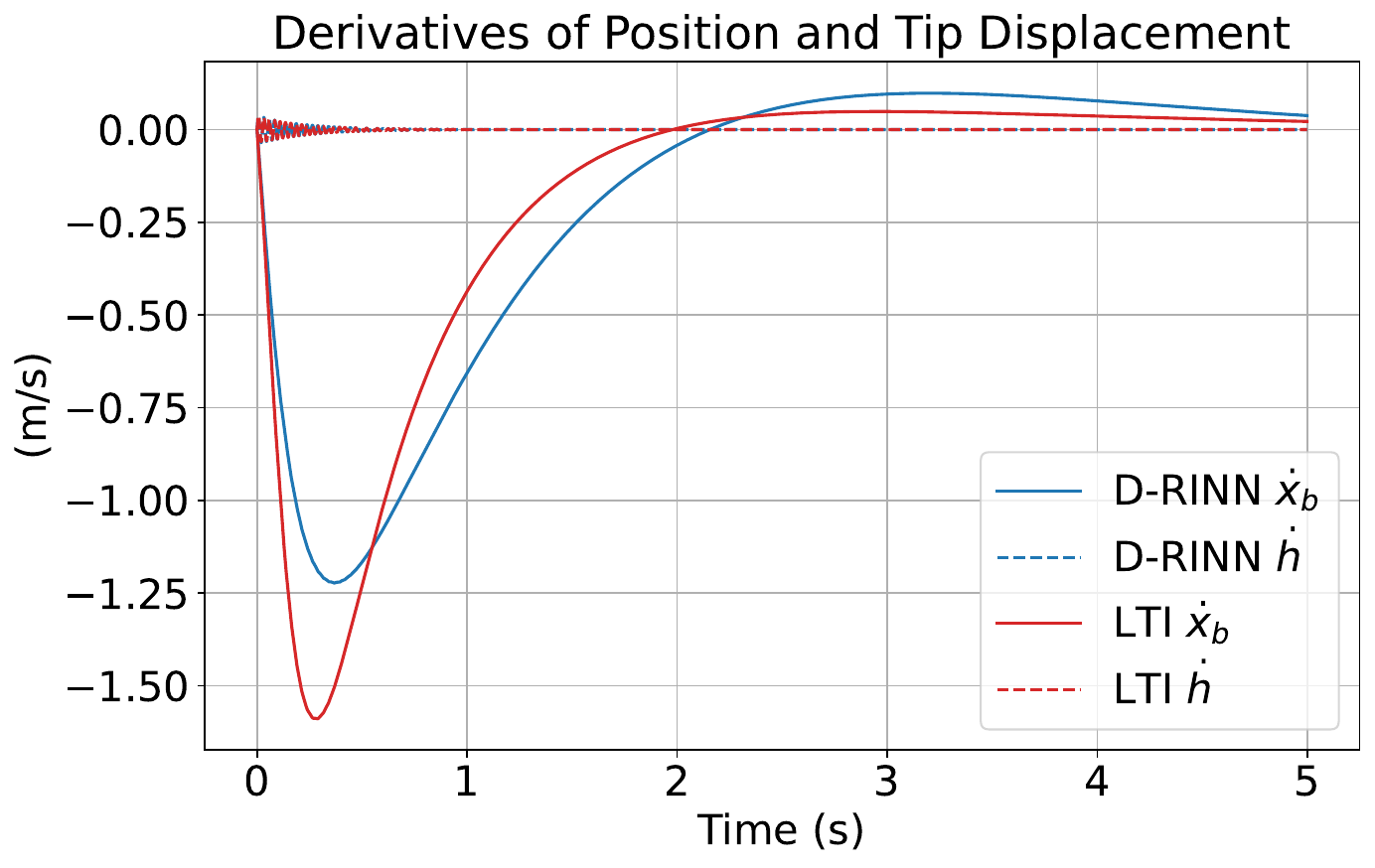}
		\label{subfig:flex-arm-time-simulation-derivatives}
	\end{subfigure}
	\hfill
	\begin{subfigure}[b]{0.3\textwidth}
		\centering
		\includegraphics[width=\textwidth]{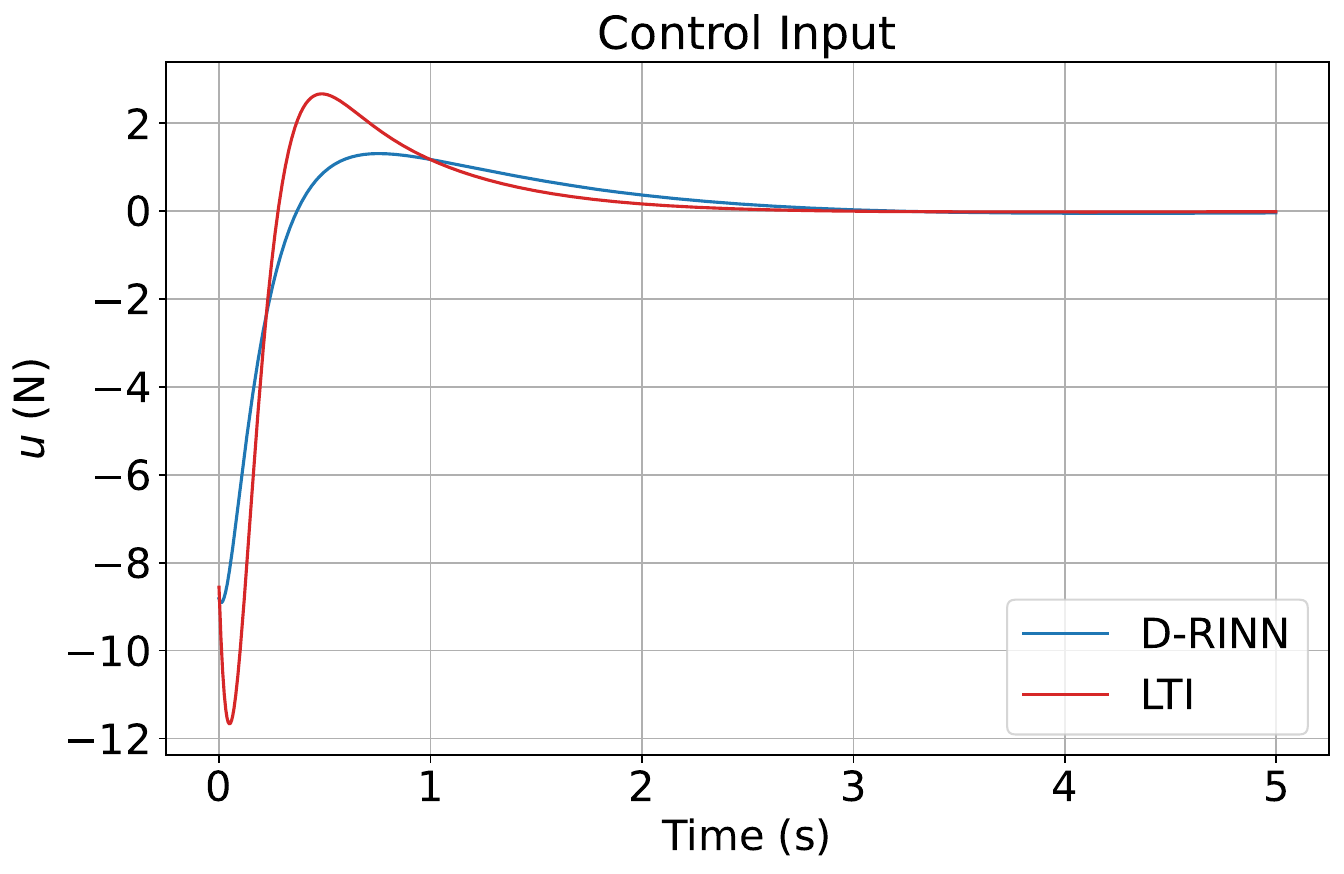}
		\label{subfig:flex-arm-time-simulation-action}
	\end{subfigure}
	\caption{
		Time simulations for the Dissipative Recurrent Implicit Neural Network (D-RINN, ours) and LTI controller starting from initial condition $\mathbf{x_f}(0)=\begin{bsmallmatrix}1 & 0 & 0 & 0\end{bsmallmatrix}^\top$ on the flexible rod on a cart.
		The D-RINN uses less control effort than the LTI controller to stabilize the system while satisfying the same robustness guarantees.
	}
	\label{fig:flex-arm-time-simulation}
\end{figure*}

The norm-bound on $\Delta$ is characterized with a quadratic constraint of $M_{\Delta_p} = 5 \begin{bsmallmatrix}
		0.1^2 & 0 \\ 0 & -1
	\end{bsmallmatrix}$.
The scaling factor of 5 is the $\lambda$ value used in Lemma~\ref{lemma:static-iqc-dissipativity}.
This affects the feasible set, but not the norm-bound.
The supply rate in synthesizing these controllers is set to $0.5 \begin{bsmallmatrix}
		0.99 & 0 \\ 0 & -I
	\end{bsmallmatrix}$, constraining the $L_2$ gain from disturbance into the control input to the rigid model state being less than or equal to 0.99.
Scaling the supply rate does not change the $L_2$ gain property.
We use a $t_{RS}$ value of 1.0 (in \eqref{eq:trs-def}) and a backoff factor of $\beta = 1.05$ to tune suboptimality of the projection (as described in Section~\ref{subsec:training}).

We simulate the controllers and plant with a time step of $0.001$ seconds, and train with a time horizon of 2 seconds.
The reward for each time step is $\exp(-\|x_f\|^2) + \exp(-u^2)$.
Initial conditions are sampled uniformly with $x_b \in [-1, 1]$ m, $h \in [-0.44, 0.44]$ m, $\dot{x}_b \in [-0.25, 0.25]$ m/s, and $\dot{h} \in [-2, 2]$ m/s.
The control input is limited (saturated) to remain in the interval $[-20, 20]$ N.

The Dissipative RINN has state size $n_k=2$ and activation function size $n_\phi=16$.
The controller is initialized with an LTI controller as described in Section~\ref{subsec:inv-pend}.
We use a learning rate of $10^{-5}$ for this model to improve stability during training.
We compare the Dissipative RINN with the following:
\begin{itemize}
	\item LTI: This model has the same state size as the Dissipative RINN, is initialized using the same method,
	      and is trained as described in Section~\ref{subsec:inv-pend}.
	      We use a learning rate of $5 \cdot 10^{-5}$ for this model.

	\item Standard RINN: This model has the same state and activation function sizes as the Dissipative RINN.
	      After each training step, $\|D_{kvw}\|_\infty$ is reduced to be less than 1 to guarantee well-posedness  \citep{el_ghaoui_implicit_2021}. We use a learning rate of $5 \cdot 10^{-5}$ for this model.

	\item Fully Connected Neural Network:
	      The fully connected neural network has two hidden layers, both with size 19, to give a total number of parameters roughly the same as the two RINN controllers.
	      We use a learning rate of $5 \cdot 10^{-5}$ for this model.

	\item Soft D-RINN: This model has the same state and activation function sizes as the Dissipative RINN and is trained with a learning rate of $5 \cdot 10^{-5}$.
	      Training follows Section~\ref{subsec:inv-pend}, with $\lambda_s = 1.0$ selected as the smallest weight for which all three seeds result in trained controllers that satisfy the dissipativity condition.

\end{itemize}

Each controller model is trained three times with different seeds.
Figure~\ref{fig:flexarm} plots the training episode reward vs. the number of training environment steps sampled.
As expected, the two unconstrained neural networks, the FCNN and the Standard RINN, achieve the highest reward.
The Dissipative RINN achieves reward close to its unconstrained counterpart and significantly higher reward than the LTI controller, while maintaining the same robustness guarantee as the LTI controller.
The gap between the fully connected neural network and the Standard RINN suggests either training difficulties with the RINN model, or the disadvantage of excluding biases from the neural network.
The Soft D-RINN fails to learn on this example, achieving approximately one fourth the episode reward of the Dissipative RINN, while both satisfy the dissipativity condition at convergence.
Smaller weights $\lambda_s$ achieve higher reward but fail the dissipativity check at convergence.
The Soft D-RINN parameters remain near the LTI initialization throughout training, indicating the strong penalty prevents the controller from improving reward beyond the initialization.

Figure~\ref{fig:flex-arm-time-simulation} shows time simulations for the best controllers of the LTI and D-RINN models with the initial state of the flexible rod on a cart being $\mathbf{x_f}(0)=\begin{bsmallmatrix}1 & 0 & 0 & 0\end{bsmallmatrix}^\top$.
Note that the time simulations are taken without saturation on the control input (these saturations are present during training), although we anyway observe that neither control input signals exceed the saturation thresholds.
The time simulations show that the Dissipative RINN, our method, uses lower control effort than the LTI controller, while converging the state to the origin in a similar time.
We do not include the unconstrained neural network controllers in the time simulation because, due to the reward function heavily incentivizing small control effort, they do not stabilize the system.

\section{Conclusion} \label{sec:conclusion}

In this paper, we derived a convex condition for certifying dissipativity of a feedback system of a neural network controller and a plant modeled by the interconnection of an LTI system, uncertainty, and nonlinearity.
The derivation of this condition leverages implicit neural networks and integral quadratic constraints to model the neural network controller and the plant in the same form.
We used this condition in a method to train neural network controllers that provide closed-loop dissipativity guarantees to maximize reward.
We applied this method to an inverted pendulum model to train a stabilizing controller to use minimal control effort, and to a flexible rod on a cart model to train a controller guaranteeing a closed-loop $L_2$ gain upper bound to stabilize the system with small control effort.
These experiments showed that the neural network controller with formal closed-loop guarantees achieved higher performance than that of an LTI controller with the same guarantees.

Future directions of research include reducing conservatism of the over-approximation of the neural network activation function, improving the scalability of this LMI-based approach, improving the numerical stability of training RINNs, and accelerating training of RINNs.


\bibliographystyle{plainnat}        
\bibliography{refs}           



\appendix
\section{Expansion of Extended and Transformed System} \label{appendix:tilde-expansion}
This section includes the expansion of the system that is the result of augmenting an uncertain LTI system with the filters corresponding to the dynamic IQC as described in Section~\ref{subsec:dynamic-iqc}.
The definitions of the tilde variables are as follows.
\begin{equation*}
	\begin{gathered}
		\begin{bmatrix}
			\tilde{A}   & \tilde{B}_w     & \tilde{B}_d     \\
			\tilde{C}_v & \tilde{D}_{v w} & \tilde{D}_{v d} \\
			\tilde{C}_e & \tilde{D}_{e w} & \tilde{D}_{e d}
		\end{bmatrix} = \\ \left[\begin{array}{ccc|c|c}
				A              & 0          & 0          & B_w                & B_d                \\
				B_{\psi_1} C_v & A_{\psi_1} & 0          & B_{\psi_1} D_{v w} & B_{\psi_1} D_{v d} \\
				0              & 0          & A_{\psi_2} & B_{\psi_2}         & 0                  \\
				\hline
				D_{\psi_1} C_v & C_{\psi_1} & 0          & D_{\psi_1} D_{v w} & D_{\psi_1} D_{v d} \\
				\hline
				C_e            & 0          & 0          & D_{e w}            & D_{e d}
			\end{array}\right]
	\end{gathered}
\end{equation*}

\section{Expansion of Closed-Loop System} \label{appendix:closed-loop-expansion}
This section includes the expansion of the feedback system of plant and controller, described in Section~\ref{sec:controller-synthesis}.
\begin{align*}
	A       & = \begin{bmatrix}
		            A_p + B_{p u} D_{k u y} C_{p y} & B_{p u} C_{k u} \\
		            B_{k y} C_{p y}                 & A_k
	            \end{bmatrix}               \\
	B_w     & = \begin{bmatrix}
		            B_{p w} + B_{p u} D_{k u y} D_{p y w} & B_{p u} D_{k u w} \\
		            B_{k y} D_{p y w}                     & B_{k w}
	            \end{bmatrix}       \\
	B_d     & = \begin{bmatrix}
		            B_{p d} + B_{p u} D_{k u y} D_{p y d} \\
		            B_{k y} D_{p y d}
	            \end{bmatrix}                           \\
	C_v     & = \begin{bmatrix}
		            C_{p v} + D_{p v u} D_{k u y} C_{p y} & D_{p v u} C_{k u} \\
		            D_{k v y} C_{p y}                     & C_{k v}
	            \end{bmatrix}       \\
	D_{v w} & = \begin{bmatrix}
		            D_{p v w} + D_{p v u} D_{k u y} D_{p y w} & D_{p v u} D_{k u w} \\
		            D_{k v y} D_{p y w}                       & D_{k v w}
	            \end{bmatrix} \\
	D_{v d} & = \begin{bmatrix}
		            D_{p v d} + D_{p v u} D_{k u y} D_{p y d} \\
		            D_{k v y} D_{p y d}
	            \end{bmatrix}                       \\
	C_e     & = \begin{bmatrix}
		            C_{p e} + D_{p e u} D_{k u y} C_{p y} & D_{p e u} C_{k u}
	            \end{bmatrix}       \\
	D_{e w} & = \begin{bmatrix}
		            D_{p e w} + D_{p e u} D_{k u y} D_{p y w} & D_{p e u} D_{k u w}
	            \end{bmatrix} \\
	D_{e d} & = D_{p e d} + D_{p e u} D_{k u y} D_{p y d}
\end{align*}


\newpage
\noindent
\begin{wrapfigure}{l}{0.9in}
	\vspace{-\baselineskip}
	\includegraphics[width=1in,height=1.25in,clip,keepaspectratio]{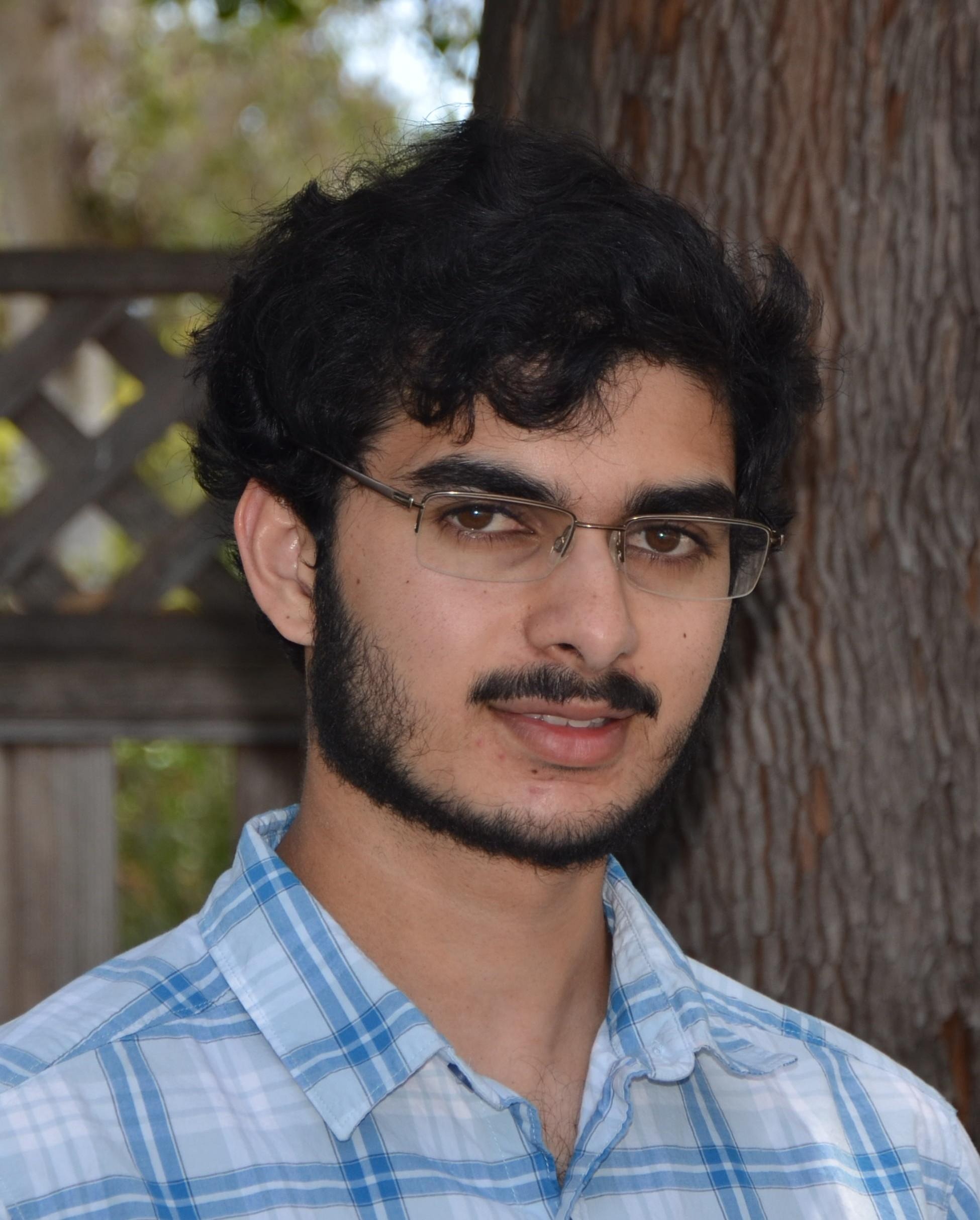}
\end{wrapfigure}
\textbf{Neelay Junnarkar} is a Ph.D. student at UC Berkeley and previously received a B.S. in Electrical Engineering and Computer Sciences from UC Berkeley (2021). His research interests include applications of machine learning to control theory.

\vspace{1em}

\noindent
\begin{wrapfigure}{l}{0.9in}
	\vspace{-\baselineskip}
	\includegraphics[width=1in,height=1.25in,clip,keepaspectratio]{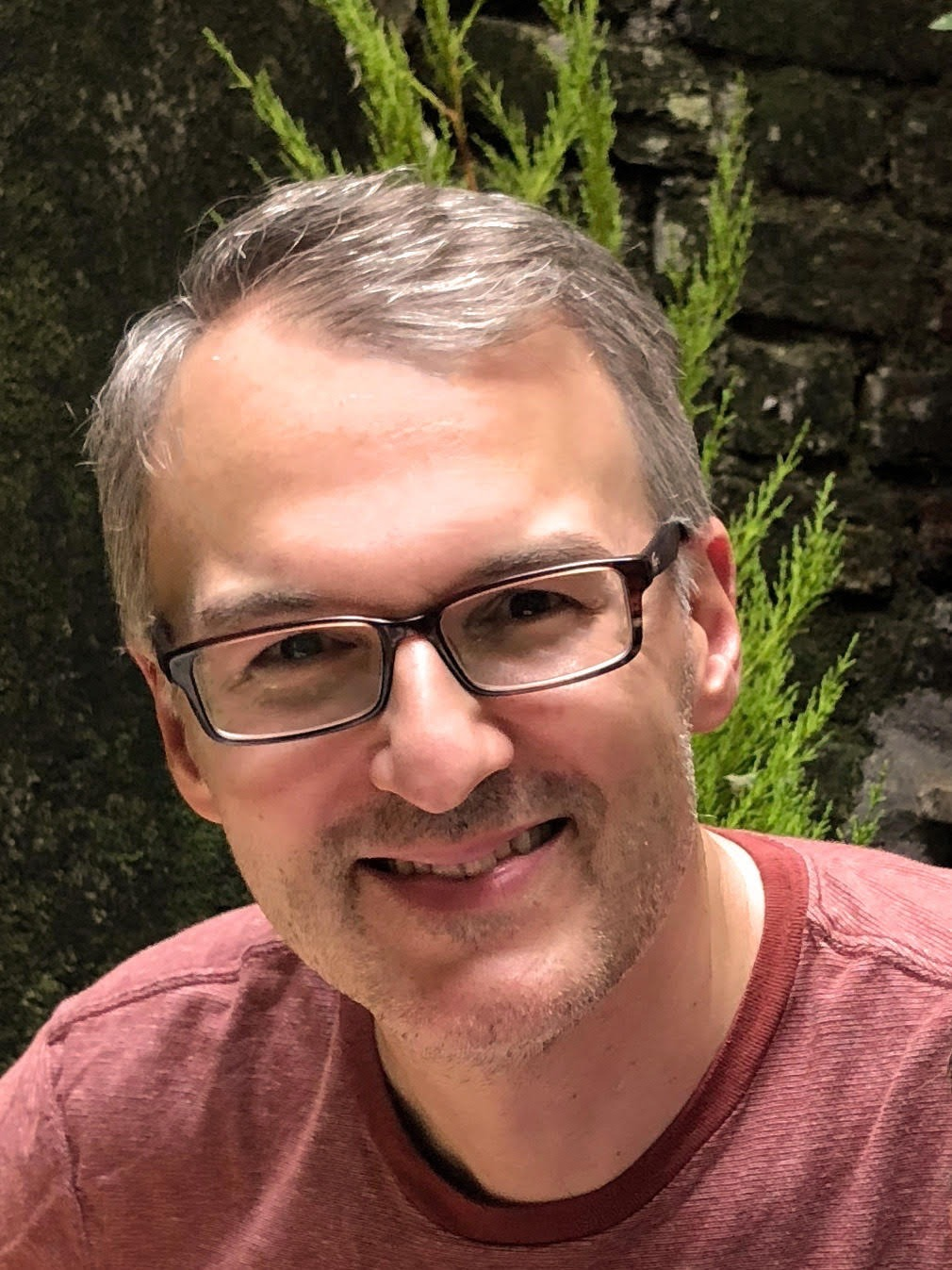}
\end{wrapfigure}
\textbf{Murat Arcak} is a professor and holds the Robert M. Saunders Endowed Chair at the University of California, Berkeley, in the Department of Electrical Engineering and Computer Sciences, with a courtesy appointment in Mechanical Engineering. He earned his B.S. degree in Electrical Engineering from Bo\u{g}azi\c{c}i University, Istanbul, in 1996, and his M.S. and Ph.D. degrees from the University of California, Santa Barbara, in 1997 and 2000, respectively. His research focuses on dynamical systems and control theory, with applications to autonomous and multi-agent systems. He has received a CAREER Award from the National Science Foundation in 2003, the Donald P. Eckman Award from the American Automatic Control Council in 2006, the Control and Systems Theory Prize from the Society for Industrial and Applied Mathematics (SIAM) in 2007, and the Antonio Ruberti Young Researcher Prize from the IEEE Control Systems Society in 2014. He is a fellow of IEEE and the International Federation of Automatic Control (IFAC).

\vspace{1em}

\noindent
\begin{wrapfigure}{l}{0.9in}
	\vspace{-\baselineskip}
	\includegraphics[width=1in,height=1.25in,clip,keepaspectratio]{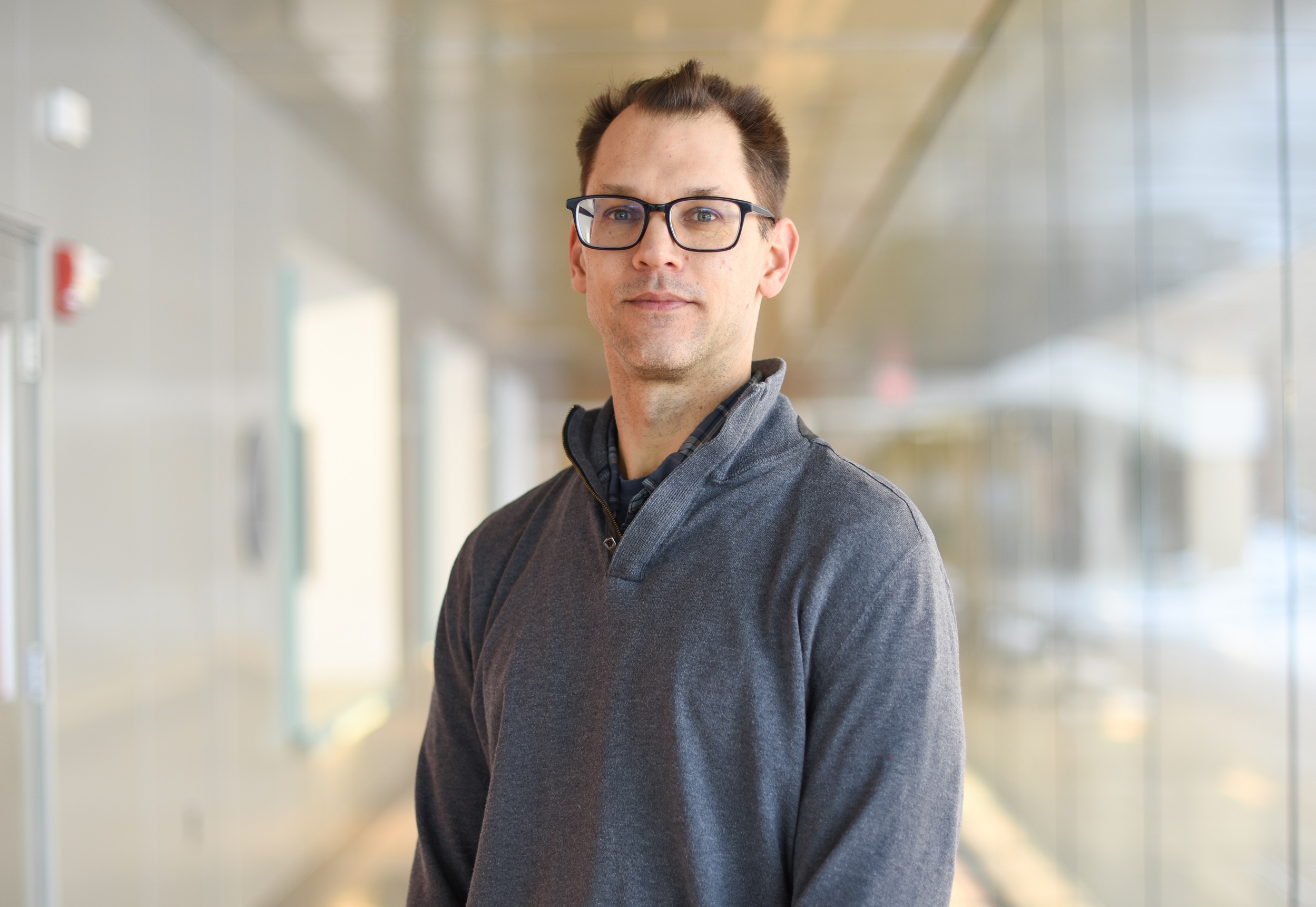}
\end{wrapfigure}
\textbf{Peter Seiler} received B.S. degrees in Mathematics and Mechanical Engineering from the University of Illinois, Urbana-Champaign in 1996. He received the Ph.D. degree in Mechanical Engineering from the University of California, Berkeley, in 2001. Since 2020, he has been a faculty member with the University of Michigan, Ann Arbor in Electrical Engineering and Computer Science. He was previously on the faculty at the University of Minnesota, Twin Cities, in Aerospace Engineering and Mechanics from 2011-2019. He was also a Principal Scientist R\&D at the Honeywell Labs in Minneapolis, Minnesota from 2004-2008. His current research interests include merging robust control techniques with online optimization and learning-based methods. He has served as an Associate Editor for Control Engineering Practice, IEEE Open Journal of Control Systems, and IEEE Control Systems Letters. Dr.~Seiler received the National Science Foundation CAREER Award in 2013, Brockett-Willems Outstanding Paper Award in 2021, and the O. Hugo Schuck Best Paper Award in 2003.

\end{document}